\documentclass[twocolumn,aps,pra,reprint,superscriptaddress,longbibliography,floatfix]{revtex4-2}
\usepackage{amsmath}
\usepackage{graphicx}
\usepackage{tabularx}
\usepackage{comment}
\usepackage{booktabs} 
\usepackage{soul}
\usepackage[table,xcdraw]{xcolor}
\usepackage{colortbl} 
\usepackage{placeins}
\usepackage{makecell}
\usepackage{subfigure}
\usepackage[most]{tcolorbox}
\usepackage[colorlinks=true,
            linkcolor=blue,
            citecolor=blue,
            urlcolor=blue]{hyperref}
\usepackage[normalem]{ulem}
\definecolor{headerblue}{rgb}{0.0, 0.2, 0.5} % Dark blue for headers
\definecolor{rowblue1}{rgb}{0.8, 0.85, 1.0}  % Light blue for alternating rows
\definecolor{rowblue2}{rgb}{0.9, 0.95, 1.0}  % Slightly lighter blue for other rows
\begin{document}
%\linenumbers
\renewcommand{\thesection}{\arabic{section}} 
\renewcommand{\thefigure}{\arabic{figure}} 
\renewcommand{\thetable}{\arabic{table}}   

\newcommand{\CoNbSe}{Co$_{1/4}$NbSe$_2$}
\newcommand{\NbSe}{NbSe$_2$}
\newcommand{\cm}{cm$^{-1}$}
\newcommand{\etg}{$E_{2g}$}
\newcommand{\aog}{$A_{1g}$}
\newcommand{\eog}{$E_{1g}$}
\newcommand{\BT}[1]{\textcolor{teal}{BT: #1}}
\newcommand{\IM}[1]{\textcolor{magenta}{IM: #1}}
\newcommand{\PV}[1]{\textcolor{blue}{PV: #1}}
\newcommand{\DB}[1]{\textcolor{olive}{DB: #1}}
\newcommand{\NJG}[1]{\textcolor{purple}{NG: #1}}
\newcommand{\Reply}[1]{\textcolor{black}{#1}}

\begin{abstract}
  We investigate the influence of Co intercalation and altermagnetic order on the lattice dynamics of the layered compound Co$_{1/4}$NbSe$_2$. Polarization-resolved Raman spectroscopy, supported by density-functional theory, enables identification of six Raman-active phonons. Co intercalation drives a substantial reconstruction of the vibrational spectrum through zone folding of NbSe$_2$ phonons, producing hybridized modes with mixed zone-center and zone-boundary character. Despite this, Co atoms do not participate in any Raman-active modes by symmetry, which is in marked contrast to related $1/3$ compounds where intercalant modes do contribute to the Raman spectrum. Temperature-dependent Raman measurements across the altermagnetic transition show no discontinuities, which is consistent with magnetic fluctuations persisting above the  N\'eel temperature. However, we find evidence for spin-phonon coupling in $A_{1g}$ symmetry modes owing to their out-of-plane Se displacements. Our work demonstrates the substantial impact of intercalation on the vibrational properties of transition metal dichalcogenides and the presence of spin-phonon interactions in a newly discovered altermagnetic material.
\end{abstract}

% \linenumbers
\title{Raman spectroscopy of the van der Waals altermagnet \texorpdfstring{Co$_{1/4}$NbSe$_2$}{Co1/4NbSe2}}

\author{Dushyanthini Balasundaram}
\thanks{These two authors contributed equally.}
\affiliation{Department of Physics and Astronomy, George Mason University, Fairfax, VA 22030, USA}
\affiliation{Quantum Science and Engineering Center, George Mason University, Fairfax, VA 22030, USA}

\author{Bishal Thapa}
\thanks{These two authors contributed equally.}
\affiliation{Department of Physics and Astronomy, George Mason University, Fairfax, VA 22030, USA}
\affiliation{Quantum Science and Engineering Center, George Mason University, Fairfax, VA 22030, USA}

\author{Resham Babu Regmi}
\affiliation{Department of Physics and Astronomy, University of Notre Dame, Notre Dame, IN 46556 USA}
\affiliation{Stavropoulos Center for Complex Quantum Matter, University of Notre Dame, Notre Dame, IN 46556 USA}

\author{Nirmal J. Ghimire}
\affiliation{Department of Physics and Astronomy, University of Notre Dame, Notre Dame, IN 46556 USA}
\affiliation{Stavropoulos Center for Complex Quantum Matter, University of Notre Dame, Notre Dame, IN 46556 USA}

\author{Igor I.\ Mazin}
\affiliation{Department of Physics and Astronomy, George Mason University, Fairfax, VA 22030, USA}
\affiliation{Quantum Science and Engineering Center, George Mason University, Fairfax, VA 22030, USA}

\author{Patrick M. Vora}% corresponding author
\email{pvora@gmu.edu}
\affiliation{Department of Physics and Astronomy, George Mason University, Fairfax, VA 22030, USA}
\affiliation{Quantum Science and Engineering Center, George Mason University, Fairfax, VA 22030, USA}
\email{pvora@gmu.edu}
\maketitle

\section{Introduction}
Intercalated transition metal dichalcogenides (TMDs) provide a versatile platform for achieving tailored magnetic phases. Monolayers of V, Cr, Mn, Fe, Co and Ni can order within the van der Waals (vdW) gap of TaS$_2$, TaSe$_2$, \NbSe{}, and NbS$_2$, resulting in ferromagnetism (FM), antiferromagnetism (AFM), chiral helimagnetism, noncoplanar AFM, and altermagnetism (AM) \cite{PhysRevX.12.040501,doi:10.1021/jacs.2c02885,PhysRevB.109.L060403,doi:10.1021/jacs.1c12975,PhysRevLett.108.107202, doi:10.1021/acs.nanolett.8b01546,5n6h-1wp3, Nair2020}. Systems hosting AM order have a FM-like electronic band structure  (spin polarized electronic bands and strong anisotropy) but with no net magnetization due to AFM spin ordering \cite{PhysRevX.12.040501,PhysRevX.12.031042}. Hence, AM materials exhibit a compensated magnetic structure, resulting in zero net magnetization while still breaking time-reversal symmetry, which ultimately leads to an alternating non-relativistic spin splitting in momentum space \cite{liu2025different}. Potential applications of AM in spintronic devices are generating considerable excitement \cite{bai_altermagnetism_2024,song_altermagnets_2025}, where the combination of spin-polarized currents and zero net magnetization could lead to powerful low-energy, high-frequency computing technologies. Numerous materials, most notably MnTe \cite{PhysRevLett.132.036702} and CrSb\cite{reimers_direct_2024}, are proven AMs \cite{PhysRevX.12.040501}. However, magnetically-intercalated TMDs provide a novel and uniquely tunable platform for achieving AM behavior \cite{regmi_altermagnetism_2025}. Recently, AM has been discovered in the intercalated TMD Co$_{{1}/{4}}$NbSe$_2$ \cite{regmi_altermagnetism_2025, dale2024nonrelativisticspinsplittingfermi, sakhya2025electronic}, and many other compounds have been predicted to be AMs \cite{dayroberts2026altermagneticmaterialslibraryintercalated} that simultaneously host novel electronic topological features such as flat bands and Weyl fermions \cite{sah_altermagnetism_2026}.

While the impact of AM on the electronic and magnetic properties of materials has been intensively studied, there is comparatively little work on the influence of AM on TMD lattice modes. Thus far, magnetically-intercalated 2H TMDs have been explored with Raman spectroscopy for both the 1/3 and 1/4 doping \cite{Musfeldt2025,PhysRevB.109.L060403,Abdullah2026,doi:10.1021/acs.nanolett.0c03292,Xie2025,Musfeldt2026_2D,Li2025}. These works confirm the existence of new modes including a prominent $A_1$ ($A_{1g}$) symmetry mode for 1/3 (1/4) doping between 100 – 200 \cm{}, depending on the intercalant \cite{doi:10.1021/acs.nanolett.0c03292,Musfeldt2025,Abdullah2026}. The tripling (quadrupling) of the original unit cell will give rise to new Raman modes because of zone-folding, which brings zone-boundary modes to the $\Gamma$ point. However, there is still substantial confusion about the mechanism through which these new peaks acquire spectral weight. Some assume that boundary modes are weak and therefore the dominant low-energy $A_1(A_{1g})$ mode must be related to the magnetic intercalants. This logic is consistent with the fact that intercalation does not change the size of the vdW gap, suggesting that the bonding between the intercalant and TMD is weak. However, it is well established that the chalcogen-intercalant distance is smaller than the vdW gap in the parent 2H TMD which suggests strong bonding. The Raman modes in 1/3 compounds do not include out-of-plane intercalant motion \cite{Abdullah2026} and in the 1/4 compound any intercalant Raman modes are forbidden by symmetry. This has not prevented many authors from labeling the $A_1(A_{1g})$ phonons ``metal-monolayer modes'' \cite{doi:10.1021/acs.nanolett.0c03292,Musfeldt2025,APL2026_FeTaSe2}. While such designation is incorrect, the mechanism behind the strength of zone-boundary modes remains unclear. 

Here we explore the impact  of intercalation on the vibrational properties of \CoNbSe{}, a recently discovered AM \cite{regmi_altermagnetism_2025}, using temperature-dependent Raman spectroscopy and density functional theory (DFT). We find that, even though Co modes are not Raman active, there is still a dramatic modification of the spectrum by the combination of zone-folding due to the 2$\times$2 commensurate Co superlattice and the subsequent hybridization between former zone-boundary modes and $\Gamma$ point modes. Polarization dependent measurements enable the identification of mode symmetry and comparison with DFT is favorable. While there is no large discontinuity at N\'eel temperature ($T_N = 168$ K) \cite{regmi_altermagnetism_2025}, we find evidence for spin-phonon coupling in $A_{1g}$ symmetry modes. This is likely due to the out-of-plane vibrational pattern of the Se atoms. Our work shows the significant impact of intercalation on the vibrational properties of TMDs and identifies the presence of AM-induced spin-phonon coupling. Both results are relevant for the broader family of 1/4 magnetically intercalated TMDs, several of which have recently been found to be AM candidates \cite{regmi_altermagnetism_2025, sah_altermagnetism_2026, dayroberts2026altermagneticmaterialslibraryintercalated}.

\section{Methods}

\subsection{Experimental Details}

Single crystals of \CoNbSe\  were grown by chemical vapor transport using iodine as the transport agent in a single zone horizontal tube furnace. Initially, a polycrystalline sample was prepared by heating stoichiometric amounts of cobalt powder (Alfa Aesar 99.998\%), niobium powder (Alfa Aesar 99.8\%), and selenium pieces (Alfa Aesar 99.9995\%) in an evacuated silica ampule at $950\,^{\circ}\mathrm{C}$ for 5 days. Subsequently, 2 g of the powder was loaded together with 0.3 g of iodine in a fused silica tube of 14 mm inner diameter. The tube was evacuated and sealed under vacuum. The ampule of 10 cm length was loaded in a horizontal tube furnace in which the temperature of the hot zone (center) was kept at $900\,^{\circ}\mathrm{C}$ for 14 days. This yielded several distinct, well-faceted \CoNbSe\ single crystals with a flat plate-like morphology. The crystals were then mechanically exfoliated in air and transferred onto a p-type silicon wafer with a 285 nm SiO$_2$ layer. Since \NbSe{} is known to oxidize rapidly in  ambient conditions \cite{doi:10.1021/acs.nanolett.5b00648,doi:10.1021/acsnano.6b08036} we restrict our attention to bulk crystals where this effect is limited.

\subsection{DFT calculations}

The structural and magnetic calculations of \CoNbSe\ were performed using the Vienna ab initio simulation package (VASP)\cite{VASP-1,VASP-2,VASP-3} and the symmetry analysis was performed with the aid of the ISOTROPY software suite \cite{stokessmodes,FINDSYM} and the Bilbao Crystallographic Server \cite{aroyo2006bilbao-1,aroyo2006bilbao-2}.
The projector augmented wave (PAW) \cite{PAW} potentials with the generalized gradient approximation (GGA) exchange-correlation potential in the Perdew–Burke–Ernzerhof \cite{PBE} as well as r2SCAN \cite{furness2020accurate} version  were used in all calculations. 
The energy cutoff in VASP was $E_{\rm cutoff}\ \approx $ 600\ eV and $8 \times 8 \times 4$  k-point sampling was used in the calculations. 
The spin-orbit interaction was not included in the calculations, as its effect on lattice properties in these materials is rather weak. The calculated magnetic moment on Co is 1.3~$\mu_{\mathrm{B}}$ for a structure based on the experimental crystallography of Ref.~\cite{regmi_altermagnetism_2025}, with the lattice parameters fixed to the experimental values and only the internal atomic coordinates relaxed. 
After a subsequent full structural relaxation, in which both the lattice parameters and the internal coordinates are optimized, the moment converges to 1.2~$\mu_{\mathrm{B}}$.

\subsection{Raman Spectroscopy}

Samples were loaded into a closed cycle helium cryostat for temperature dependent Raman spectroscopy. We measured spectra between 5 and 300~K in a backscattering geometry using a 532~nm excitation laser focused through a 40\(\times\) objective (NA = 0.6). The incident laser power was maintained at 300~\(\mu\)W (measured before the objective), and each spectrum was collected with an integration time of 600~s. Scattered photons were collected through the same objective and directed to an imaging spectrograph equipped with a LN2-cooled charge-coupled-device. Rayleigh scattering was suppressed using three Bragg filters (Optigrate), 
providing access to Raman shifts down to 10~cm$^{-1}$. For polarization-resolved measurements an analyzer is inserted between the objective and the spectrometer.

\begin{figure}[b]
    \centering
 \includegraphics[height= 10cm, width=0.50\textwidth]{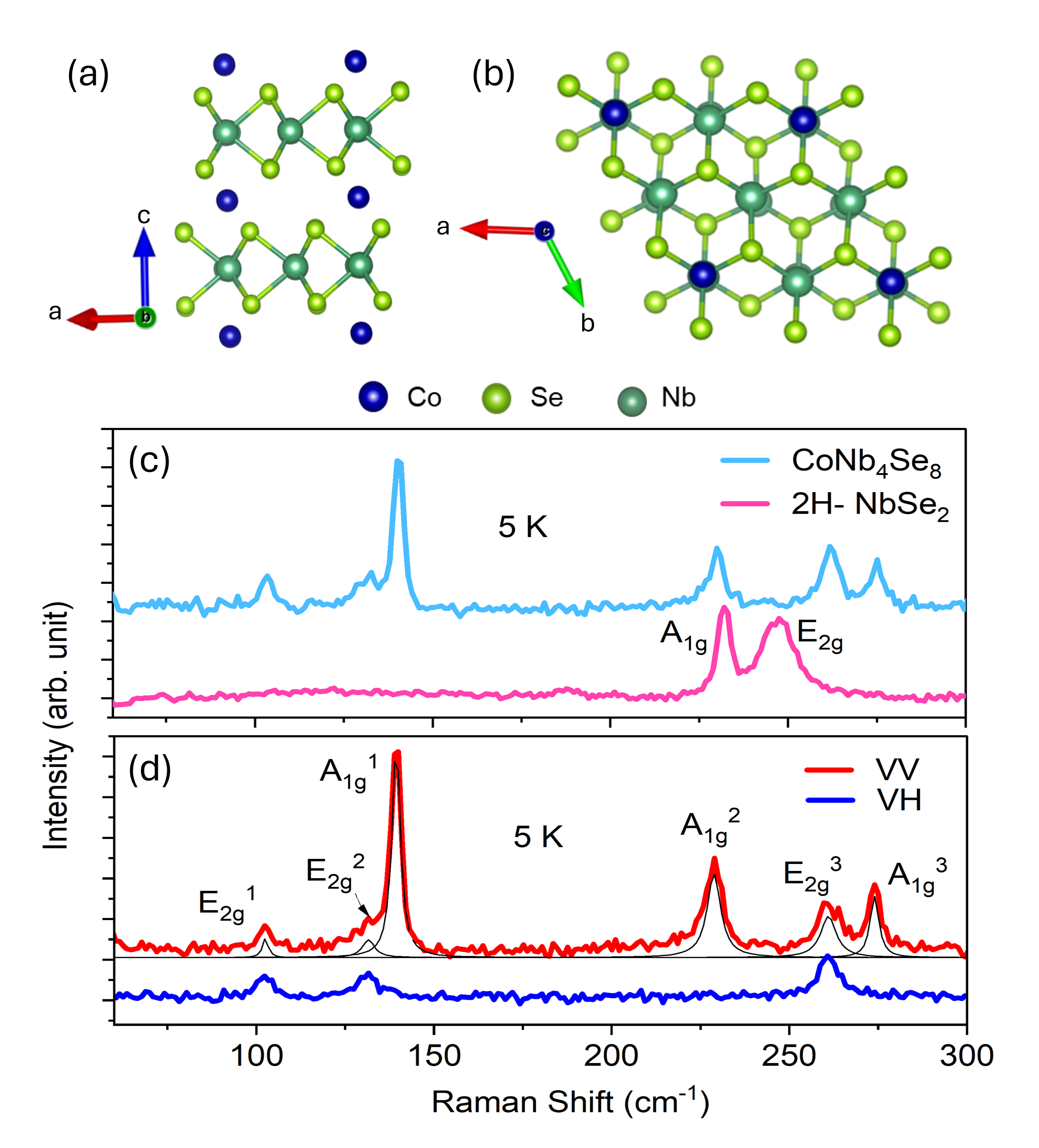}
    \caption{Lattice structure of \CoNbSe{} viewed along the (a) $b$ axis and (b) $c$ axis. (c) 5 K unpolarized Raman spectra for \CoNbSe{} and \NbSe{} illustrating the impact of Co intercalation. (d) 5 K polarized Raman spectra for \CoNbSe{} in the VV and VH configurations. These data facilitate the identification of mode symmetry as $E_{2g}$ modes are visible in both VV and VH, but $A_{1g}$ is only active in VV.}
    \label{F1}
\end{figure}

\section{Results}

We first discuss the structure of the parent compound \NbSe{} before examining \CoNbSe{}. 2H-\NbSe{} features a layered structure in which niobium (Nb) atoms are coordinated by selenium (Se) atoms in the trigonal-prismatic geometry.  The unit cell contains two distinct, symmetry-equivalent Nb atomic positions, forming two separate layers. In the lower layer, the Nb atoms are positioned at fractional coordinates \((1/3, 2/3, 1/4)\), while in the upper layer the Nb atoms are similarly positioned at \((2/3, 1/3, 1/4)\). Se atoms are located above and below each Nb atom which completes the structure. These alternating layers are stacked along the $c$-axis, resulting in a structure with space group symmetry $P6_3/mmc$, with room-temperature hexagonal lattice parameters
$a=b=3.45$~\AA{} and $c=12.54$~\AA{}\cite{regmi_altermagnetism_2025, aroyo2006bilbao-2,brown1965} and held together by weak vdW forces.

Upon intercalation, Co atoms coordinate with 1/4 of the Nb atoms, fitting in the vdW gap between the layers and creating a commensurate 2$\times$2 superlattice as shown in Fig.~\ref{F1}(a,b). Both calculation\textcolor{magenta}{s} and experiment indicate that the interlayer distance is essentially unchanged \cite{regmi_altermagnetism_2025}, with single-crystal X-ray diffraction on \CoNbSe{} (space group $P6_3/mmc$, $T=295$~K) gives $a=b=6.9178(13)$~\AA{} and $c=12.390(3)$~\AA{}, and Wyckoff positions Co on $2a$, Nb on $2b$ and $6h$, and Se on $4f$ and $12k$.
The Co superlattice doubles the unit cell in each direction and therefore drives zone-folding in reciprocal space, bringing zone boundary modes to the $\Gamma$ point.

In Fig.~\ref{F1}(c) we compare the Raman spectra of \NbSe{} and \CoNbSe{}, both taken at 5 K which is well below the charge density wave \cite{xi_strongly_2015} and N\'{e}el temperature $T_N \approx 168$~K \cite{regmi_altermagnetism_2025}, respectively. The parent compound NbSe$_2$ exhibits four Raman-active modes, $2E_{2g}+E_{1g}+A_{1g}$, corresponding to seven modes when the twofold degeneracy of the $E$ modes is explicitly counted. At 300 K we observe an \aog{} mode at 228 \cm{}, \etg{} mode at 237 \cm{}, the $E_{2g}$ interlayer shear mode at 28~cm$^{-1}$, and the well-known two-phonon mode at 176 cm$^{-1}$ (Fig. S10) \cite{shi_enhanced_2026,xi_strongly_2015,PhysRevB.98.165109}. At 5 K (Fig. \ref{F1}c) only the high frequency $A_{1g}$ and $E_{2g}$ modes are observed. The two-phonon mode has weakened due to the charge density wave transition and the interlayer shear mode is weak at low temperature under 532 nm excitation \cite{he_interlayer_2016} but can be observed by a resonant effect using 633 nm light \cite{PhysRevB.98.165109,shi_enhanced_2026,xi_strongly_2015}. \eog{} modes are not visible in our experimental geometry, a fact we will discuss later using Raman tensor analysis.

The Raman spectrum of \CoNbSe{} is quite distinct from \NbSe{} despite Co intercalation not changing the size of the vdW gap \cite{regmi_altermagnetism_2025}. Symmetry considerations predict the presence of 16 Raman-active modes, $4A_{1g}+5E_{1g}+7E_{2g}$, corresponding to 28 modes when the twofold degeneracy of the $E$ modes is explicitly counted. The additional Raman-active modes arise from the Nb $6h$ and Se $12k$ Wyckoff positions introduced by the $2\times2$ Co superlattice. Of these, we detect six well-resolved Raman peaks, only one of which appears to be close in energy to the \aog{} peak in \NbSe. Polarization resolved measurements allow for the assignment of the mode symmetry for this space group which are \aog{}, \eog{}, and \etg{} Raman modes. The corresponding Raman tensors are \cite{aroyo2006bilbao-2, YuCardona2010}:

\begin{equation}
R_{A_{1g}}:  \begin{pmatrix}
a & 0 & 0 \\
0 & a & 0 \\
0 & 0 & b
\end{pmatrix}
\end{equation}

\begin{equation}
R_{E_{1g}}:
\begin{pmatrix}
0 & 0 & 0 \\
0 & 0 & c \\
0 & c & 0
\end{pmatrix},
\begin{pmatrix}
0 & 0 & c \\
0 & 0 & 0 \\
c & 0 & 0
\end{pmatrix}
\end{equation}

\begin{equation}
R_{E_{2g}}: \begin{pmatrix}
d & 0 & 0 \\
0 & -d & 0 \\
0 & 0 & 0
\end{pmatrix}, 
\begin{pmatrix}
0 & d & 0 \\
d & 0 & 0 \\
0 & 0 & 0
\end{pmatrix}
\end{equation}

The Raman intensity $I$ of a mode can then be computed from the tensors following
\begin{equation}
I = |e_s^{T} \cdot R \cdot e_i|^2,
\end{equation}

where $e_i$ ($e_s$) are the polarization states of the incident (scattered) light fields. The experimental geometry is such that excitation and detection occur along the $c$ axis of the crystal (\textit{i.e.} a backscattering geometry), implying that the laser polarization lies within the basal plane of \CoNbSe{}. Vertical polarization (V), defined in the laboratory frame, is denoted as $\begin{pmatrix} 1 & 0 & 0 \end{pmatrix}$ and the horizontal polarization (H)  vector is $\begin{pmatrix} 0 & 1 & 0 \end{pmatrix}$. For the co-polarized case (VV), where the laser and the analyzer are both V, the Raman intensities are $I_{A{_{1g}}}^{VV} = \lvert a \rvert^2,\text{ }I_{E{_{1g}}}^{VV} = 0,\text{ and }I_{E{_{2g}}}^{VV} = \lvert d \rvert^2$, implying both the \aog{} and \etg{} modes are detectable while the \eog{} mode is not. Repeating the exercise for the cross-polarized (VH) case where the laser is V and the analyzer is H, we find $I_{A{_{1g}}}^{VH} = 0, \text{ }I_{E{_{1g}}}^{VH} = 0,  \text{ and }I_{E{_{2g}}}^{VH} = \lvert d \rvert^2.$ This implies that \etg{} is visible in VV and VH, \aog{} is only visible in VV, and that \eog{} modes are not detectable in our measurements. $E_{1g}$ modes could be observed by exciting the sample with the laser polarization having a significant component along the $c$-axis.

In Fig.~\ref{F1}(d) we present polarization-resolved Raman measurements for \CoNbSe{} in both the VV and VH configurations at 5 K. By comparing these two cases we can easily isolate the \aog{} symmetry modes from the \etg{} modes based on the selection rules described above. Fitting the spectra to a sum of Lorentzian functions enables extraction of the mode frequencies which are reported in Table~S9 along with the corresponding mode symmetry. We observe three \aog{} and three \etg{} Raman modes, corresponding to six well-resolved Raman peaks.

The impact of Co intercalation on the Raman modes of NbSe$_2$ is significant, as evidenced by density functional theory (DFT) calculations. The intercalated Co atoms form a $2\times2$ superlattice that is commensurate with the Nb atoms, leading to the folding of zone-boundary (Z.B.) phonon modes to the $\Gamma$ point of the reduced Brillouin zone. Upon folding, these Z.B. modes strongly hybridize with the intrinsic $\Gamma$-point modes, giving rise to the spectral features observed in Fig.~1(c) and 1(d).

Consistent with this picture, DFT calculations performed for Co$_{1/4}$NbSe$_2$ reveal the activation of additional $\Gamma$-point modes as compared to NbSe$_2$, thereby enabling $\Gamma$--Z.B. mode hybridization. In \CoNbSe{}, the computed $\Gamma$-point branches $A_{1g}^{1}$, $A_{1g}^{2}$, $A_{1g}^{3}$ and $E_{2g}^{1}$, $E_{2g}^{2}$, $E_{2g}^{3}$ display alternating-phase Nb displacements with doubled real-space periodicity and finite Se out-of-plane components (see Supplementary Material Tables S3-S8), which are clear fingerprints of zone folding and mode mixing. Within the $A_{1g}$ family, $A_{1g}^{2}$ most closely resembles the \NbSe{} $A_{1g}(\Gamma)$ breathing mode, whereas the complementary $A_{1g}^{1}$ and $A_{1g}^{3}$ modes exhibit the strongest mixing, combining dominant Se $z$ motion with sizable alternating in-plane Nb oscillations. The low-frequency $A_{1g}^{1}$ mode, formerly a Z.B. mode and activated by zone folding, shows the highest intensity due to its strong coupling with Co sites.  For the $E_{2g}$ family, $E_{2g}^{1}$ and $E_{2g}^{2}$ are predominantly folded Z.B. $E$-type modes that gain Raman activity through symmetry lowering, whereas only $E_{2g}^{3}$ can be regarded as originating from the  $E_{2g}(\Gamma)$ mode of 2H-\NbSe{}. However, even this correspondence is approximate due to the Co-induced 2$\times$2 superlattice potential, which introduces small Se $z$ components and alternating-phase Nb displacements that hybridize $\Gamma$ and Z.B.\ modes.

Symmetry analysis shows that all Raman-active modes of \CoNbSe{} arise exclusively from displacements of Nb and Se, with no displacement of Co \cite{stokessmodes,aroyo2006bilbao-1,aroyo2006bilbao-2}.  This is in contrast to the conventional viewpoint in the literature where intercalated magnetic monolayers exhibit out-of-plane and in-plane superlattice vibrations in the frequency range of 120 - 200 \cm{} \cite{doi:10.1021/acs.nanolett.0c03292,doi:10.1021/jacs.2c02885, doi:10.1021/acs.jpcc.3c00870}. These modes would correspond to the $E_{2g}^2$ and $A_{1g}^{2}$ peaks in our measurements. However, our DFT calculations unambiguously show that for the 1/4 compound no true Co monolayer vibrational modes exist (Figs. S3-S8). The closest computed mode resembling a metal monolayer vibration is a silent $E_{2u}$ mode at 133~\cm{}, which still involves vibrations of both Nb and Se atoms. Our results therefore refute the assignment of intercalate superlattice modes, at least for 1:4:2 stoichiometry compounds.

It is reasonable to suspect that magnetic order could impact the vibrational modes of \CoNbSe{}. We have examined this via DFT calculations of \CoNbSe{} Raman modes for the non-magnetic (NM), FM, and AFM state, the latter of which has the same spin ordering as the AM state. As shown in Table~S9, all three calculations show reasonable agreement with experiment except for select modes.  NM frequencies are within 5 cm$^{-1}$ for all modes other than $E_{2g}^1$ where it exhibits a significant discrepancy. This may originate from the different effective ionic radius of Co in the AFM/FM versus NM states. The FM case gives the best overall agreement with the data, other than a large mismatch with $A_{1g}^2$, while the AFM case exhibits the worst agreement with the $A_{1g}^1$ mode being off by 17 cm$^{-1}$.  We verified that these conclusions are robust with respect to the choice of exchange-correlation functional (GGA-PBE and meta-GGA r$^2$SCAN; see Methods). Although the FM calculations give phonon frequencies in closer agreement with experiment than the AFM calculations, we find no obvious physical mechanism for this apparent improvement and therefore regard it as a fortuitous cancellation of errors in the FM calculation, rather than evidence that the FM state provides a more appropriate description of the low-temperature magnetic state.

\begin{figure}[t]
    \centering
    \includegraphics[height= 9cm,width=0.48\textwidth]{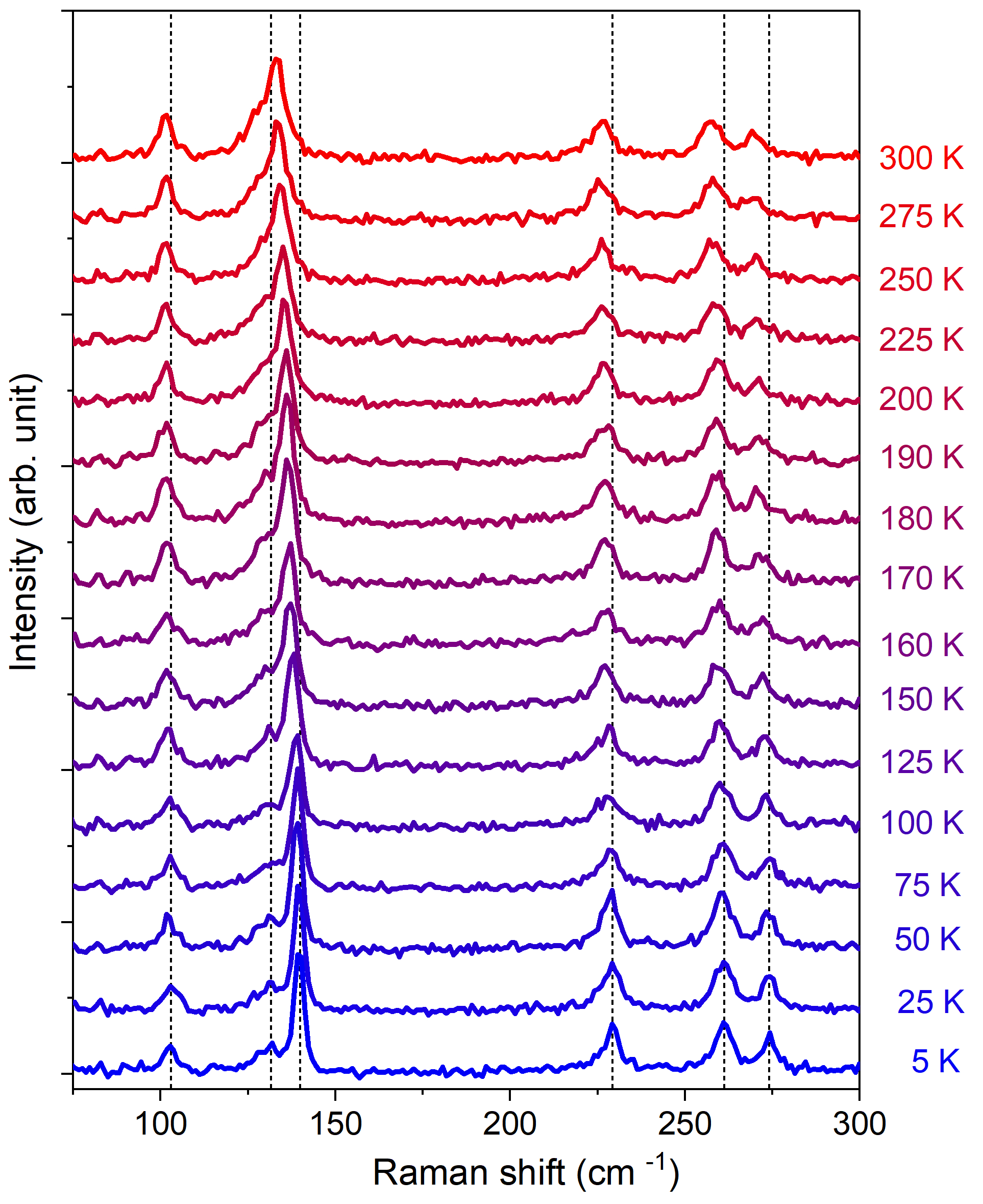}
    \caption{Temperature evolution of Raman spectra from 5 K to 300 K, offset for clarity.}
    \label{Fig:tempdep}
\end{figure}

To understand the impact of the AM transition on vibrational modes, we perform temperature-dependent Raman measurements across the AM transition, which occurs at the N\'{e}el temperature $T_N \approx 168$~K. Figure~\ref{Fig:tempdep} presents the temperature evolution of the Raman spectra from 5~K to 300~K, documenting the expected redshift of all modes due to lattice expansion. By fitting each spectrum to a sum of six Lorentzian functions (as in Fig.~\ref{F1}d), we extract the temperature dependence of each mode. We average results taken on three separate bulk flakes to account for sample variations, a procedure presented in the Supplementary Material Fig. S9.

\begin{figure}[t]
     \centering
     \includegraphics[width=\columnwidth]{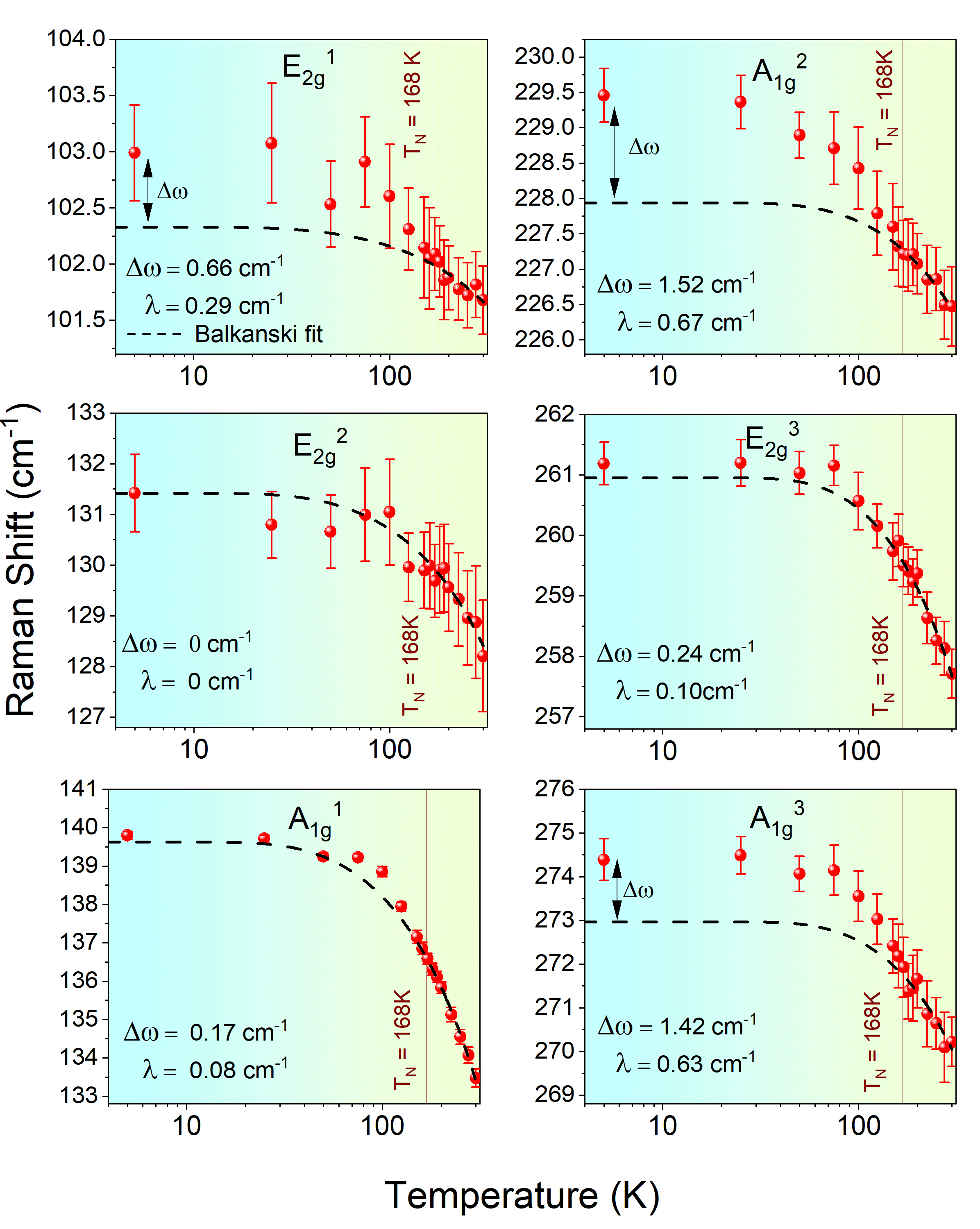}
     \caption{Temperature dependence of Raman frequency shifts for different phonon modes of Co$_{1/4}$NbSe$_2$. Dashed lines show fits using the Balkanski anharmonic model in Eq. \eqref{eq:Balkanski}. $\Delta \omega$  values are obtained as described in the text}
     \label{fig:all thermo}
 \end{figure}

Our results are shown in Fig.~\ref{fig:all thermo} on a lin-log scale with $T_N$ indicated by a vertical line. We find the Raman mode frequencies do not exhibit large discontinuities at $T_N$ in contrast to several other TMD magnetic intercalates \cite{doi:10.1063/1.4914134,doi:10.1021/acs.nanolett.0c03292}. At first glance this suggests weak spin-phonon coupling, despite the significant impact of Co intercalation on the Raman spectrum of \CoNbSe{}.

Spin-phonon coupling in magnetic materials is often determined by analyzing the deviation of $\omega(T)$ from the expected anharmonic behavior above the ordering temperature. This is achieved by fitting the high-temperature data to a model function and extrapolating it to 0~K. The difference between the measured and fit values is proportional to the spin-phonon coupling constant \cite{baltensperger_influence_1970,lockwood_spin-phonon_1988,fennie_magnetically_2006}. In our case, the lowest measured temperature is 5 K and so the relationship becomes

\begin{equation}
\Delta \omega \equiv \omega_{exp}(5 \text{ K}) - \omega_{fit}(5 \text{ K}) = \lambda \langle S_i \cdot S_j \rangle,
\label{eq:spin_phonon}
\end{equation}

where $\omega_{fit}(5 \text{ K})$ is the extrapolated value from the fit, $\omega_{exp}(5 \text{ K})$ is the measured frequency, $\lambda$ is the spin--phonon coupling constant, and $S_i$ and $S_j$ represent the spins of neighboring Co atoms\cite{baltensperger_influence_1970, lockwood_spin-phonon_1988, fennie_magnetically_2006}.

Different functions have been used to model anharmonic behaviors above the magnetic ordering temperature. Some groups use a phenomenological sigmoid--Boltzmann function~\cite{doi:10.1021/acs.nanolett.0c03292,
doi:10.1021/acs.jpcc.3c00870,doi:10.1063/1.4914134} whereas others use a model based on optical phonon decay (often referred to as the Balkanski model)~\cite{balkanski_anharmonic_1983}. We have tried both approaches and found that the sigmoid--Boltzmann function can occasionally produce poor fits, placing the inflection point and high temperature asymptote at incorrect locations. Moreover, the model itself was initially motivated to capture glass-like behavior in the paramagnetic state of CrSiTe$_3$,~\cite{doi:10.1063/1.4914134} which we do not expect in Co$_{1/4}$NbSe$_2$. Therefore we use the Balkanski model to describe anharmonic effects above $T_N$ which is given by

\begin{equation}
\omega_m(T) = \omega_{0m} + A_m \left( 1 + \frac{2}{e^{x} - 1} \right),
\label{eq:Balkanski}
\end{equation}

where $\omega_m(T)$ is the temperature-dependent frequency of mode $m$, $\omega_{0m}$ is the frequency of mode $m$ in the absence of anharmonic effects, $A_m$ is a constant representing the magnitude of the anharmonicity, and $x=\hbar\omega_{0m}/2k_BT$. This formulation describes the decay of one optical phonon into two phonons and has been used to study spin-phonon coupling in a wide range of materials \cite{PhysRevB.78.064307,PhysRevB.73.214301,Tian_2016,pandey_spinphonon_2026,bhattacharya_mean_2026}.

In Fig.~\ref{fig:all thermo} we present fits of all six Raman-active modes to Eq.\eqref{eq:Balkanski} over the  temperature range 170 K - 300 K (above $T_N$) and extrapolate those fits to 0~K. The fits capture the behavior above $T_N$ well and we find that for several Raman-active modes, there is a significant deviation from the experimental frequencies. This difference is the quantity $\Delta\omega$ defined in Eq.~\eqref{eq:spin_phonon} which allows us to determine the spin--phonon coupling constants once the value of $\langle S_i \cdot S_j \rangle$ is specified. Since magnetic fluctuations may persist above $T_N$, we tested the sensitivity of the extracted $\Delta\omega$ to the choice of fitting range by varying the lower fitting bound from 170 to 200~K while keeping the upper bound fixed at 300~K (see Supplementary Fig.~S11 and Table~S10). The resulting $\Delta\omega$ values do not change significantly across these fitting windows which suggests the impact of magnetic fluctuations is small immediately above $T_N$.

Prior neutron scattering measurements by members of our collaboration have determined that the magnetic moment is $1.34\,\mu_B$ per Co atom \cite{regmi_altermagnetism_2025} which is lower than the $\tfrac{3}{2}$ expected for a fully localized Co$^{2+}$ ion. The reduced magnetic moment obtained from neutron diffraction reflects the itinerant character of the Co $3d$ electrons and their hybridization with the NbSe$_2$ host lattice. Such reductions are commonly observed in intercalated TMDs and are consistent with DFT calculations that predict a smaller ordered moment \cite{PhysRevB.110.144420,PhysRevResearch.4.013048,VANDENBERG1968143}. Despite this itinerant behavior, the use of an effective localized spin $S_i = S_j = \tfrac{3}{2}$ provides a reasonable approximation for describing magnetic interactions within the framework of conventional spin--phonon coupling models. With this assumption, we extract the spin--phonon coupling constant $\lambda$ for each mode from $\Delta \omega /2.25$ \cite{doi:10.1063/1.4914134}.

\begin{table}[!t]
\caption{Phonon frequency shifts $\Delta\omega$ and spin-phonon coupling constants
$\lambda$ at 5 K.}
\label{tab:phonon_shifts}
\begin{ruledtabular}
\begin{tabular}{lcc}
Mode & $\Delta\omega$ (cm$^{-1}$) & $\lambda$ (cm$^{-1}$) \\
\hline
$E^{1}_{2g}$ & 0.66         & 0.29 \\
$E^{2}_{2g}$ & $\approx 0$  & $\approx 0$  \\
$A^{1}_{1g}$ & 0.17         & 0.08  \\
$A^{2}_{1g}$ & 1.52         & 0.67 \\
$E^{3}_{2g}$ & 0.24         & 0.10 \\
$A^{3}_{1g}$ & 1.42         & 0.63 \\
\end{tabular}
\end{ruledtabular}
\end{table}

We summarize the $\lambda$ values obtained by this approach in Table~1, finding that the results are highly mode dependent. The largest values are observed for the $A_{1g}^{2}$ and $A_{1g}^{3}$ modes. The $E_{2g}^{1}$ mode also seems to show signs of spin--phonon coupling; however, the error bars for the frequency are large owing to the weak intensity of this feature, and so this result may be questionable. The remaining modes show comparatively small differences between the fit and the experimental data leading to small values for $\lambda$.

The Supplementary Material reports the DFT $\Gamma$-point eigenvectors for the observed $A_{1g}$ and $E_{2g}$ modes (Tables~S3--S8; Figs.~S3--S8). As mentioned previously, for all Raman-active modes the Co atoms are stationary. Therefore, coupling can only occur indirectly between the Co and the Se atoms if the latter has significant displacement along the $c$ axis into the vdW gap. Examining the eigenvectors in the Supplementary Material we see that all $E_{2g}$ modes have predominantly in-plane character while the $A_{1g}$ modes have significant displacement of the Se atoms along the $c$ axis. For $A_{1g}^{2}$ and $A_{1g}^{3}$ the Se atoms move towards the stationary Co atoms whereas for $A_{1g}^{1}$ the Se atoms move towards the Nb layer. Therefore, the $A_{1g}^{2}$ and $A_{1g}^{3}$ modes should be sensitive to the change in the magnetic ordering of the Co atoms while the $A_{1g}^{1}$ mode is not. This intuition is consistent with our experimental results, providing a clear connection between spin-phonon coupling and atomic displacements. The presence of a small change in the behavior of $\omega(T)$ at $T_N$, rather than a large discontinuity, is consistent with magnetic fluctuations persisting above the phase transition despite the absence of long-range AM order. Unfortunately, our Raman measurements cannot determine the spatial range of these correlations.

\section{Conclusion}

In this study we have combined temperature-dependent Raman spectroscopy with DFT calculations to understand how Co intercalation and AM order modify the vibrational modes of \CoNbSe{}. The ordering of the Co atoms in a 2$\times$2 superlattice increases the number of Raman active modes through zone folding and $\Gamma$-Z.B. mode hybridization. We find that despite a common interpretation of modes in the 100 - 200 \cm{} range as arising from the magnetic atom superlattice, in 1/4 intercalation compounds all Co vibrations are inactive and these features actually arise from zone-folding and mode hybridization alone. Temperature-dependent Raman measurements clearly reveal strong spin--phonon coupling in two $A_{1g}$ modes where Se atoms are displaced toward the Co atoms. This suggests an indirect interaction between phonons and the AM superlattice which is sensitive to the phonon eigenvectors. The lack of a substantial discontinuity at $T_N$ is consistent with magnetic fluctuations persisting above the phase transition. These results show that intercalation and magnetic symmetry breaking both strongly modify the Raman spectrum of \CoNbSe{} and clarify misunderstandings about mode origin in 1/4 compounds. We expect the intuition derived from this study to be broadly applicable to other AM systems derived from intercalated TMDs.

\begin{acknowledgments}
P.M.V. and D.B. are supported by the U.S. National Science Foundation under grant no. 2226097. N.J.G. and I.I.M. were supported by  Army Research Office under Cooperative Agreement Number W911NF- 22-2-0173. 

\end{acknowledgments}

\bibliographystyle{apsrev4-2}
\bibliography{biblio}
\end{document}

\clearpage
\appendix

\renewcommand{\thetable}{S\arabic{table}}
\renewcommand{\thefigure}{S\arabic{figure}}

\newcommand{\CoNbSe}{Co$_{1/4}$NbSe$_2$}

\setcounter{table}{0}
\setcounter{figure}{0}

% \linenumbers
\title{Raman spectroscopy of the van der Waals altermagnet \texorpdfstring{Co$_{1/4}$NbSe$_2$}{Co1/4NbSe2}}

\author{Dushyanthini Balasundaram}
\thanks{These two authors contributed equally.}
\affiliation{Department of Physics and Astronomy, George Mason University, Fairfax, VA 22030, USA}
\affiliation{Quantum Science and Engineering Center, George Mason University, Fairfax, VA 22030, USA}

\author{Bishal Thapa}
\thanks{These two authors contributed equally.}
\affiliation{Department of Physics and Astronomy, George Mason University, Fairfax, VA 22030, USA}
\affiliation{Quantum Science and Engineering Center, George Mason University, Fairfax, VA 22030, USA}

\author{Resham Regmi}
\affiliation{Department of Physics and Astronomy, University of Notre Dame, Notre Dame, IN 46556 USA}
\affiliation{Stavropoulos Center for Complex Quantum Matter, University of Notre Dame, Notre Dame, IN 46556 USA}

\author{Nirmal J. Ghimire}
\affiliation{Department of Physics and Astronomy, University of Notre Dame, Notre Dame, IN 46556 USA}
\affiliation{Stavropoulos Center for Complex Quantum Matter, University of Notre Dame, Notre Dame, IN 46556 USA}

\author{Igor I.\ Mazin}
\affiliation{Department of Physics and Astronomy, George Mason University, Fairfax, VA 22030, USA}
\affiliation{Quantum Science and Engineering Center, George Mason University, Fairfax, VA 22030, USA}

\author{Patrick M. Vora}% corresponding author
\email{pvora@gmu.edu}
\affiliation{Department of Physics and Astronomy, George Mason University, Fairfax, VA 22030, USA}
\affiliation{Quantum Science and Engineering Center, George Mason University, Fairfax, VA 22030, USA}
\email{pvora@gmu.edu}
\maketitle

\section{Raman Active NbSe$_2$ Phonons}

% -------- TABLE 1 --------
\begin{minipage}[t]{0.98\textwidth}
\centering
\captionof{table}{Phonon mode at 236.92~cm$^{-1}$ ($E_{2g}$) in 2H--NbSe$_2$}
\footnotesize 
\setlength{\tabcolsep}{7pt} 
\renewcommand{\arraystretch}{1.05} 
\begin{tabular}{lccc}
\hline
Atom & $\Delta x$ & $\Delta y$ & $\Delta z$ \\
\hline
Nb1  &  0.561  &  0.000  &  0.000 \\
Nb2  & -0.561  &  0.000  &  0.000 \\
Se1  & -0.304  &  0.000  &  0.000 \\
Se2  & -0.304  &  0.000  &  0.000 \\
Se3  &  0.304  &  0.000  &  0.000 \\
Se4  &  0.304  &  0.000  &  0.000 \\
\hline
\end{tabular}
\end{minipage}

\vspace{0.4cm}

% -------- FIGURE 1 --------

\begin{minipage}[t]{0.98\textwidth}
\centering
\includegraphics[width=0.6\linewidth]{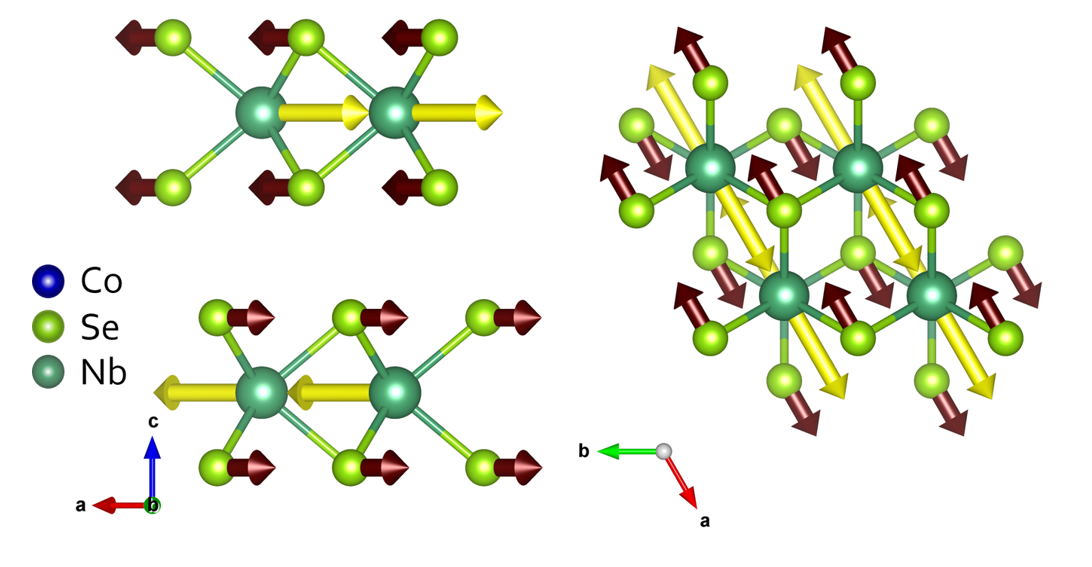}
\captionof{figure}{Two orientations of the 236.92~cm$^{-1}$ ($E_{2g}$) phonon mode in 2H--NbSe$_2$.}
\end{minipage}

\vspace{1.5cm}

% -------- TABLE 2 --------
\begin{minipage}[t]{0.98\textwidth}
\centering
\captionof{table}{Phonon mode at 215.86~cm$^{-1}$ ($A_{1g}$) in 2H--NbSe$_2$}
\footnotesize 
\setlength{\tabcolsep}{7pt} 
\renewcommand{\arraystretch}{1.05} 
\begin{tabular}{lccc}
\hline
Atom & $\Delta x$ & $\Delta y$ & $\Delta z$ \\
\hline
Nb1  &  0.000  &  0.000  &  0.000 \\
Nb2  &  0.000  &  0.000  &  0.000 \\
Se1  &  0.000  &  0.000  & -0.500 \\
Se2  &  0.000  &  0.000  &  0.500 \\
Se3  &  0.000  &  0.000  & -0.500 \\
Se4  &  0.000  &  0.000  &  0.500 \\
\hline
\end{tabular}
\end{minipage}

\vspace{0.4cm}

% -------- FIGURE 2 --------
\begin{minipage}[t]{0.98\textwidth}
\centering
\includegraphics[width=0.6\linewidth]{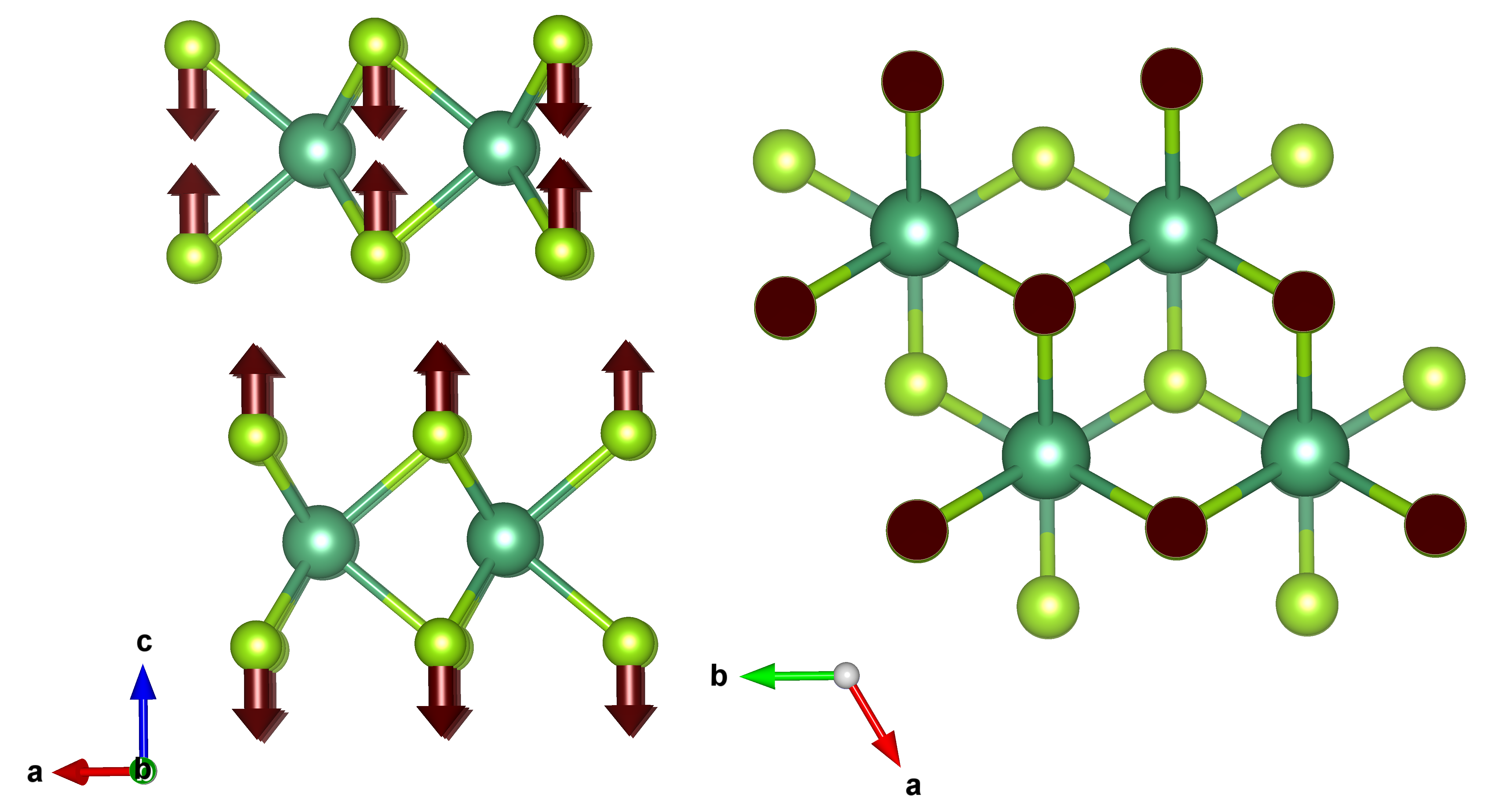}
\captionof{figure}{Two orientations of the 215.86~cm$^{-1}$ ($A_{1g}$) phonon mode in 2H--NbSe$_2$.}
\end{minipage}

% -------- TABLE 3 --------
\clearpage
\onecolumngrid
\section{DFT Calculations of Observed \CoNbSe{} Raman Modes in the AFM State}

\begin{center}
\begin{minipage}[t]{0.98\textwidth}
\centering
\captionof{table}{$E_{2g}^1$ phonon mode calculated at 98.18~cm$^{-1}$.}
\label{STab:E3}
\footnotesize
\setlength{\tabcolsep}{8pt}
\renewcommand{\arraystretch}{1.1}
\begin{tabular}{lccc}
\hline
Atom & $\Delta x$ & $\Delta y$ & $\Delta z$ \\
\hline
Co1 & -0.000 & 0.000 & 0.000 \\
Co2 & -0.000 & -0.000 & 0.000 \\
Nb1 & -0.171 & 0.017 & 0.000 \\
Nb2 & -0.334 & 0.169 & 0.000 \\
Nb3 & 0.171 & -0.017 & 0.000 \\
Nb4 & 0.334 & -0.169 & -0.000 \\
Nb5 & 0.037 & -0.009 & -0.000 \\
Nb6 & -0.037 & 0.009 & 0.000 \\
Nb7 & -0.098 & 0.129 & 0.000 \\
Nb8 & 0.098 & -0.129 & -0.000 \\
Se1 & 0.063 & 0.104 & -0.119 \\
Se2 & -0.111 & 0.056 & 0.000 \\
Se3 & 0.063 & 0.088 & -0.099 \\
Se4 & -0.111 & 0.056 & 0.000 \\
Se5 & 0.111 & -0.056 & -0.000 \\
Se6 & -0.063 & -0.088 & 0.099 \\
Se7 & -0.063 & -0.104 & 0.119 \\
Se8 & 0.166 & -0.100 & 0.218 \\
Se9 & 0.111 & -0.056 & -0.000 \\
Se10 & -0.166 & 0.100 & -0.218 \\
Se11 & -0.063 & -0.104 & -0.119 \\
Se12 & 0.063 & 0.104 & 0.119 \\
Se13 & -0.063 & -0.088 & -0.099 \\
Se14 & 0.063 & 0.088 & 0.099 \\
Se15 & -0.166 & 0.100 & 0.218 \\
Se16 & 0.166 & -0.100 & -0.218 \\
\hline
\end{tabular}
\end{minipage}

\vspace{1.0cm}

% ---- Figure Below ----
\begin{minipage}[t]{0.85\textwidth}
\centering
\includegraphics[width=0.95\linewidth]{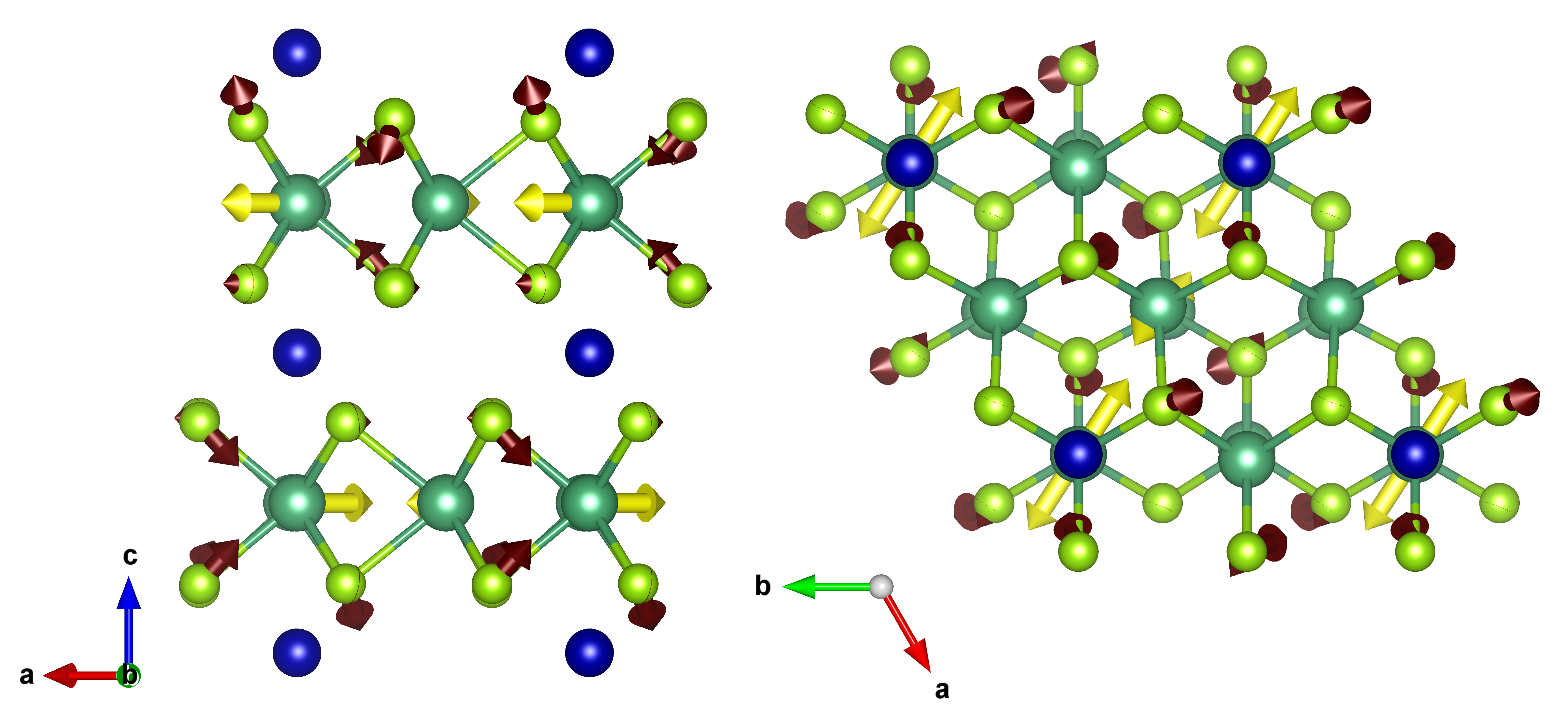}
\captionof{figure}{Two orientations of the 98.18~cm$^{-1}$ ($E^1_{2g}$) phonon mode in Co$_{1/4}$NbSe$_2$.
Left: view along the $b$-axis; right: view along the $c$-axis.
Arrows denote atomic displacement directions.}
\end{minipage}
\end{center}
\clearpage

\begin{center}
\begin{minipage}[t]{0.98\textwidth}
\centering
\captionof{table}{$E_{2g}^2$ phonon mode calculated at 124.17~cm$^{-1}$.}
\label{STab:E2}
\footnotesize
\setlength{\tabcolsep}{8pt}
\renewcommand{\arraystretch}{1.1}
\begin{tabular}{lccc}
\hline
Atom & $\Delta x$ & $\Delta y$ & $\Delta z$ \\
\hline
Co1 & -0.000 & 0.000 & 0.000 \\
Co2 & 0.000 & 0.000 & 0.000 \\
Nb1 & 0.200 & -0.174 & 0.000 \\
Nb2 & -0.016 & -0.016 & -0.000 \\
Nb3 & -0.200 & 0.174 & -0.000 \\
Nb4 & 0.016 & 0.016 & -0.000 \\
Nb5 & -0.086 & -0.270 & -0.000 \\
Nb6 & 0.086 & 0.270 & -0.000 \\
Nb7 & -0.242 & -0.050 & 0.000 \\
Nb8 & 0.242 & 0.050 & 0.000 \\
Se1 & 0.151 & 0.074 & 0.159 \\
Se2 & -0.163 & -0.167 & -0.000 \\
Se3 & 0.031 & -0.125 & -0.118 \\
Se4 & -0.163 & -0.167 & 0.000 \\
Se5 & 0.163 & 0.167 & 0.000 \\
Se6 & -0.031 & 0.125 & 0.118 \\
Se7 & -0.151 & -0.074 & -0.159 \\
Se8 & 0.016 & -0.058 & -0.041 \\
Se9 & 0.163 & 0.167 & -0.000 \\
Se10 & -0.016 & 0.058 & 0.041 \\
Se11 & -0.151 & -0.074 & 0.159 \\
Se12 & 0.151 & 0.074 & -0.159 \\
Se13 & -0.031 & 0.125 & -0.118 \\
Se14 & 0.031 & -0.125 & 0.118 \\
Se15 & -0.016 & 0.058 & -0.041 \\
Se16 & 0.016 & -0.058 & 0.041 \\
\hline
\end{tabular}
\end{minipage}

\vspace{1.0cm}

% ---- Figure Below ----
\begin{minipage}[t]{0.85\textwidth}
\centering
\includegraphics[width=0.95\linewidth]{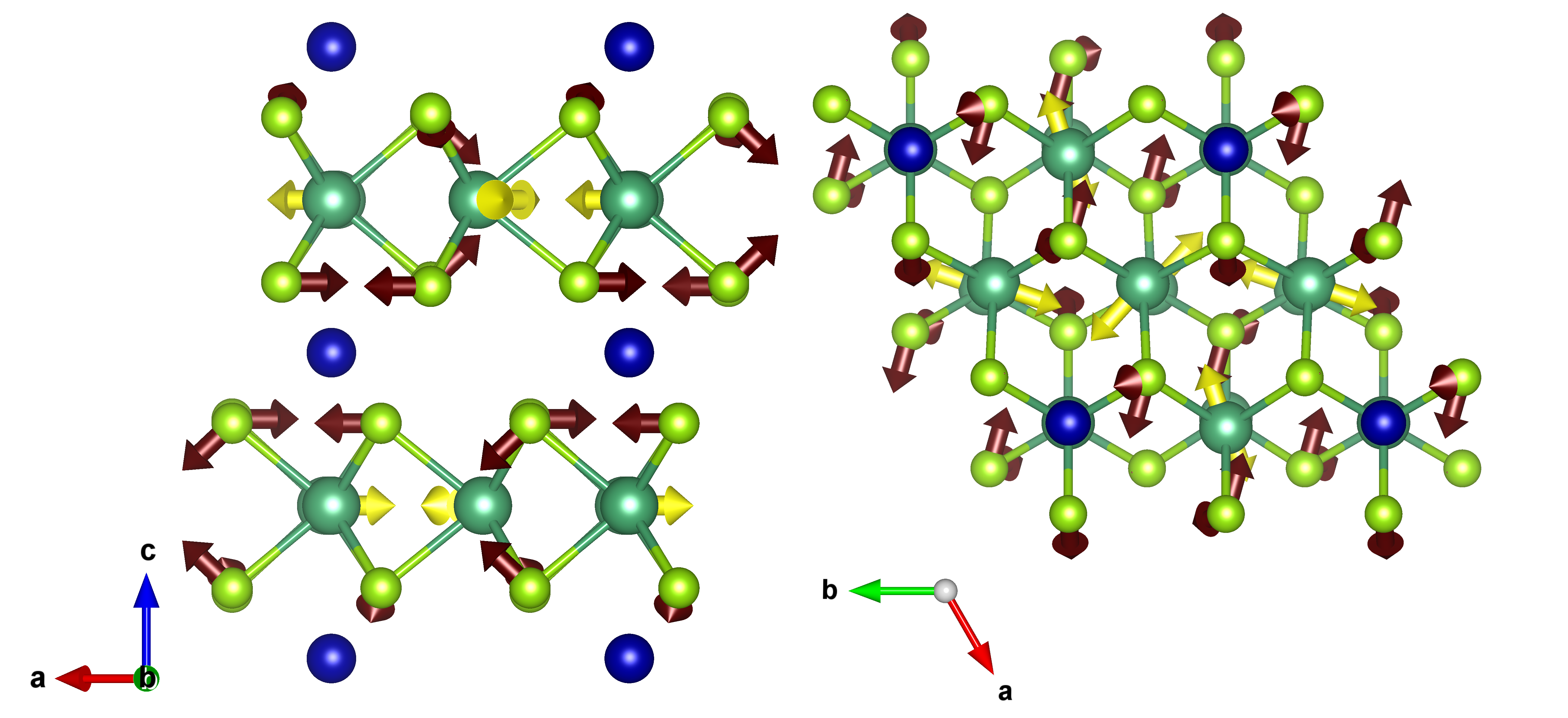}
\captionof{figure}{Two orientations of the 124.17~cm$^{-1}$ ($E^2_{2g}$) phonon mode in Co$_{1/4}$NbSe$_2$.
Left: view along the $b$-axis; right: view along the $c$-axis.
Arrows denote atomic displacement directions.}
\end{minipage}
\end{center}
\clearpage

\begin{center}
\begin{minipage}[t]{0.98\textwidth}
\centering
\captionof{table}{$A_{1g}^1$ phonon mode calculated at 150.35~cm$^{-1}$.}
\label{STab:A3}
\footnotesize
\setlength{\tabcolsep}{8pt}
\renewcommand{\arraystretch}{1.1}
% \begin{tabular}{lccc}
% \hline
% Atom & $\Delta x$ & $\Delta y$ & $\Delta z$ \\
% \hline
% Co1 & -0.000 & 0.000 & -0.000 \\
% Co2 & 0.000 & -0.000 & -0.000 \\
% Nb1 & -0.000 & 0.259 & -0.000 \\
% Nb2 & 0.000 & -0.000 & 0.000 \\
% Nb3 & 0.000 & -0.259 & -0.000 \\
% Nb4 & -0.000 & 0.000 & 0.000 \\
% Nb5 & -0.224 & 0.130 & 0.000 \\
% Nb6 & 0.224 & -0.130 & 0.000 \\
% Nb7 & -0.224 & -0.130 & -0.000 \\
% Nb8 & 0.224 & 0.130 & -0.000 \\
% Se1 & 0.017 & 0.010 & -0.102 \\
% Se2 & 0.000 & 0.000 & -0.342 \\
% Se3 & -0.000 & 0.019 & -0.102 \\
% Se4 & 0.000 & 0.000 & 0.342 \\
% Se5 & -0.000 & 0.000 & 0.342 \\
% Se6 & 0.000 & -0.019 & 0.102 \\
% Se7 & -0.017 & -0.010 & 0.102 \\
% Se8 & -0.017 & 0.010 & -0.102 \\
% Se9 & 0.000 & -0.000 & -0.342 \\
% Se10 & 0.017 & -0.010 & 0.102 \\
% Se11 & -0.017 & -0.010 & -0.102 \\
% Se12 & 0.017 & 0.010 & 0.102 \\
% Se13 & -0.000 & -0.019 & -0.102 \\
% Se14 & 0.000 & 0.019 & 0.102 \\
% Se15 & 0.017 & -0.010 & -0.102 \\
% Se16 & -0.017 & 0.010 & 0.102 \\
% \hline
% \end{tabular}
\begin{tabular}{lccc}
\hline
Atom & $\Delta x$ & $\Delta y$ & $\Delta z$ \\
\hline
Co1  & -0.000 &  0.000 & -0.000 \\
Co2  &  0.000 & -0.000 & -0.000 \\
Nb1  & -0.000 &  0.259 & -0.000 \\
Nb2  &  0.000 & -0.000 &  0.000 \\
Nb3  &  0.000 & -0.259 & -0.000 \\
Nb4  & -0.000 &  0.000 &  0.000 \\
Nb5  & -0.224 &  0.130 &  0.000 \\
Nb6  &  0.224 & -0.130 &  0.000 \\
Nb7  & -0.224 & -0.130 & -0.000 \\
Nb8  &  0.224 &  0.130 & -0.000 \\
Se1  &  0.017 &  0.010 & -0.102 \\
Se2  &  0.000 &  0.000 & -0.342 \\
Se3  & -0.000 &  0.019 & -0.102 \\
Se4  &  0.000 &  0.000 &  0.342 \\
Se5  & -0.000 &  0.000 &  0.342 \\
Se6  &  0.000 & -0.019 &  0.102 \\
Se7  & -0.017 & -0.010 &  0.102 \\
Se8  & -0.017 &  0.010 & -0.102 \\
Se9  &  0.000 & -0.000 & -0.342 \\
Se10 &  0.017 & -0.010 &  0.102 \\
Se11 & -0.017 & -0.010 & -0.102 \\
Se12 &  0.017 &  0.010 &  0.102 \\
Se13 & -0.000 & -0.019 & -0.102 \\
Se14 &  0.000 &  0.019 &  0.102 \\
Se15 &  0.017 & -0.010 & -0.102 \\
Se16 & -0.017 &  0.010 &  0.102 \\
\hline
\end{tabular}

\end{minipage}

\vspace{1.0cm}

% ---- Figure Below ----
\begin{minipage}[t]{0.85\textwidth}
\centering
\includegraphics[width=0.95\linewidth]{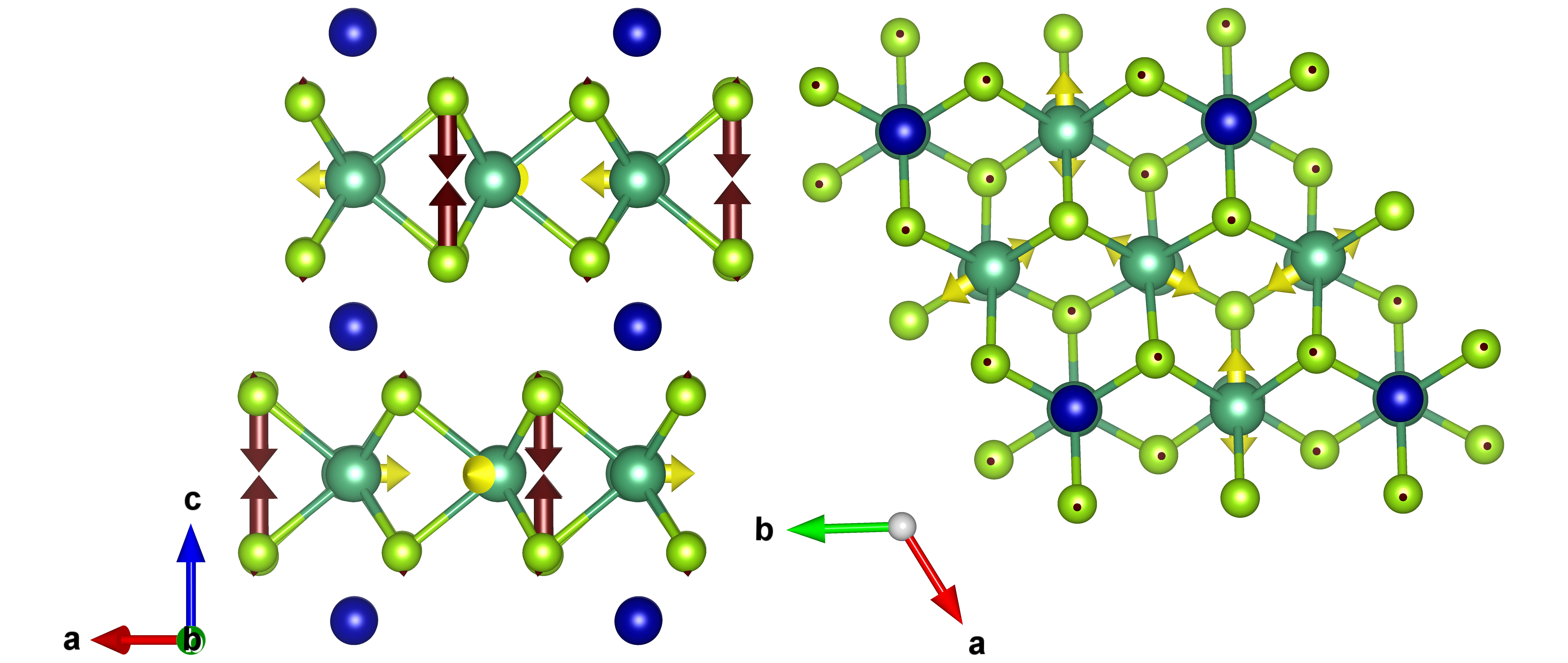}
\captionof{figure}{Two orientations of the 150.35~cm$^{-1}$ ($A^1_{1g}$) phonon mode in Co$_{1/4}$NbSe$_2$.
Left: view along the $b$-axis; right: view along the $c$-axis.
Arrows denote atomic displacement directions.}
\end{minipage}
\end{center}
\clearpage

\begin{center}
\begin{minipage}[t]{0.98\textwidth}
\centering
\captionof{table}{$A_{1g}^2$ phonon mode calculated at 220.36~cm$^{-1}$.}
\label{STab:A2}
\footnotesize
\setlength{\tabcolsep}{8pt}
\renewcommand{\arraystretch}{1.1}
\begin{tabular}{lccc}
\hline
Atom & $\Delta x$ & $\Delta y$ & $\Delta z$ \\
\hline
Co1 & -0.000 & 0.000 & 0.000 \\
Co2 & 0.000 & -0.000 & -0.000 \\
Nb1 & -0.000 & -0.046 & 0.000 \\
Nb2 & 0.000 & -0.000 & 0.000 \\
Nb3 & 0.000 & 0.046 & -0.000 \\
Nb4 & 0.000 & -0.000 & -0.000 \\
Nb5 & 0.040 & -0.023 & -0.000 \\
Nb6 & -0.040 & 0.023 & -0.000 \\
Nb7 & 0.040 & 0.023 & -0.000 \\
Nb8 & -0.040 & -0.023 & 0.000 \\
Se1 & -0.036 & -0.021 & -0.265 \\
Se2 & 0.000 & 0.000 & 0.177 \\
Se3 & 0.000 & -0.041 & -0.265 \\
Se4 & 0.000 & -0.000 & -0.177 \\
Se5 & 0.000 & -0.000 & -0.177 \\
Se6 & -0.000 & 0.041 & 0.265 \\
Se7 & 0.036 & 0.021 & 0.265 \\
Se8 & 0.036 & -0.021 & -0.265 \\
Se9 & 0.000 & -0.000 & 0.177 \\
Se10 & -0.036 & 0.021 & 0.265 \\
Se11 & 0.036 & 0.021 & -0.265 \\
Se12 & -0.036 & -0.021 & 0.265 \\
Se13 & -0.000 & 0.041 & -0.265 \\
Se14 & -0.000 & -0.041 & 0.265 \\
Se15 & -0.036 & 0.021 & -0.265 \\
Se16 & 0.036 & -0.021 & 0.265 \\
\hline
\end{tabular}
\end{minipage}

\vspace{1.0cm}

% ---- Figure Below ----
\begin{minipage}[t]{0.85\textwidth} % slightly narrower for balance
\centering
\includegraphics[width=0.95\linewidth]{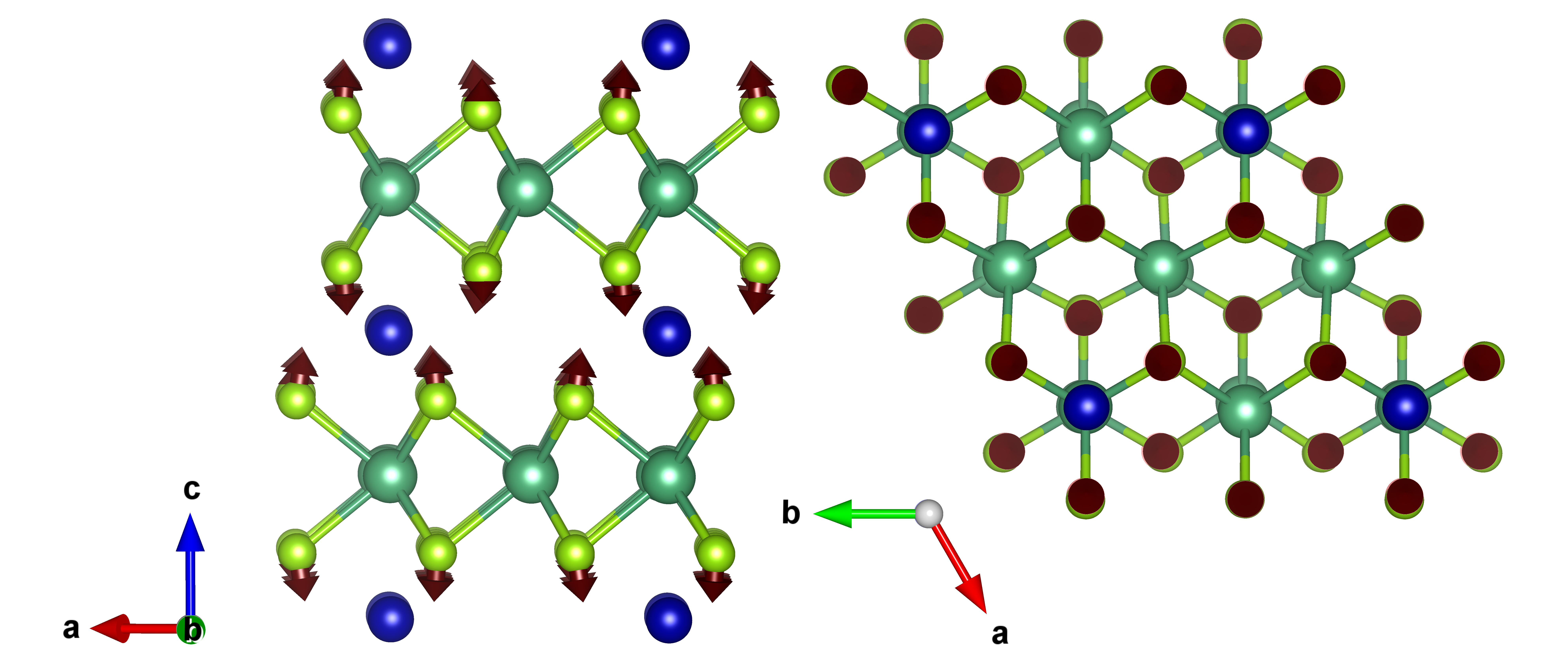}
\captionof{figure}{Two orientations of the 220.36~cm$^{-1}$ ($A^2_{1g}$) phonon mode in Co$_{1/4}$NbSe$_2$.
Left: view along the $b$-axis; right: view along the $c$-axis.
Arrows denote atomic displacement directions.}
\end{minipage}
\end{center}
\clearpage

\begin{center}
\begin{minipage}[t]{0.98\textwidth}
\centering
\captionof{table}{$E_{2g}^3$ phonon mode calculated at 259.12~cm$^{-1}$.}
\label{STab:E1}
\footnotesize
\setlength{\tabcolsep}{8pt}
\renewcommand{\arraystretch}{1.1}
\begin{tabular}{lccc}
\hline
Atom & $\Delta x$ & $\Delta y$ & $\Delta z$ \\
\hline
Co1 & 0.000 & 0.000 & -0.000 \\
Co2 & 0.000 & 0.000 & -0.000 \\
Nb1 & -0.260 & -0.059 & -0.000 \\
Nb2 & -0.307 & -0.079 & -0.000 \\
Nb3 & 0.260 & 0.059 & 0.000 \\
Nb4 & 0.307 & 0.079 & 0.000 \\
Nb5 & 0.242 & 0.077 & -0.000 \\
Nb6 & -0.242 & -0.077 & -0.000 \\
Nb7 & -0.235 & -0.053 & 0.000 \\
Nb8 & 0.235 & 0.053 & 0.000 \\
Se1 & -0.010 & -0.045 & 0.202 \\
Se2 & 0.102 & 0.026 & 0.000 \\
Se3 & -0.124 & 0.008 & -0.052 \\
Se4 & 0.102 & 0.026 & 0.000 \\
Se5 & -0.102 & -0.026 & -0.000 \\
Se6 & 0.124 & -0.008 & 0.052 \\
Se7 & 0.010 & 0.045 & -0.202 \\
Se8 & 0.025 & 0.089 & -0.150 \\
Se9 & -0.102 & -0.026 & -0.000 \\
Se10 & -0.025 & -0.089 & 0.150 \\
Se11 & 0.010 & 0.045 & 0.202 \\
Se12 & -0.010 & -0.045 & -0.202 \\
Se13 & 0.124 & -0.008 & -0.052 \\
Se14 & -0.124 & 0.008 & 0.052 \\
Se15 & -0.025 & -0.089 & -0.150 \\
Se16 & 0.025 & 0.089 & 0.150 \\
\hline
\end{tabular}
\end{minipage}

\vspace{1.0cm}

% ---- Figure Below ----
\begin{minipage}[t]{0.85\textwidth} % slightly narrower for balance
\centering
\includegraphics[width=0.95\linewidth]{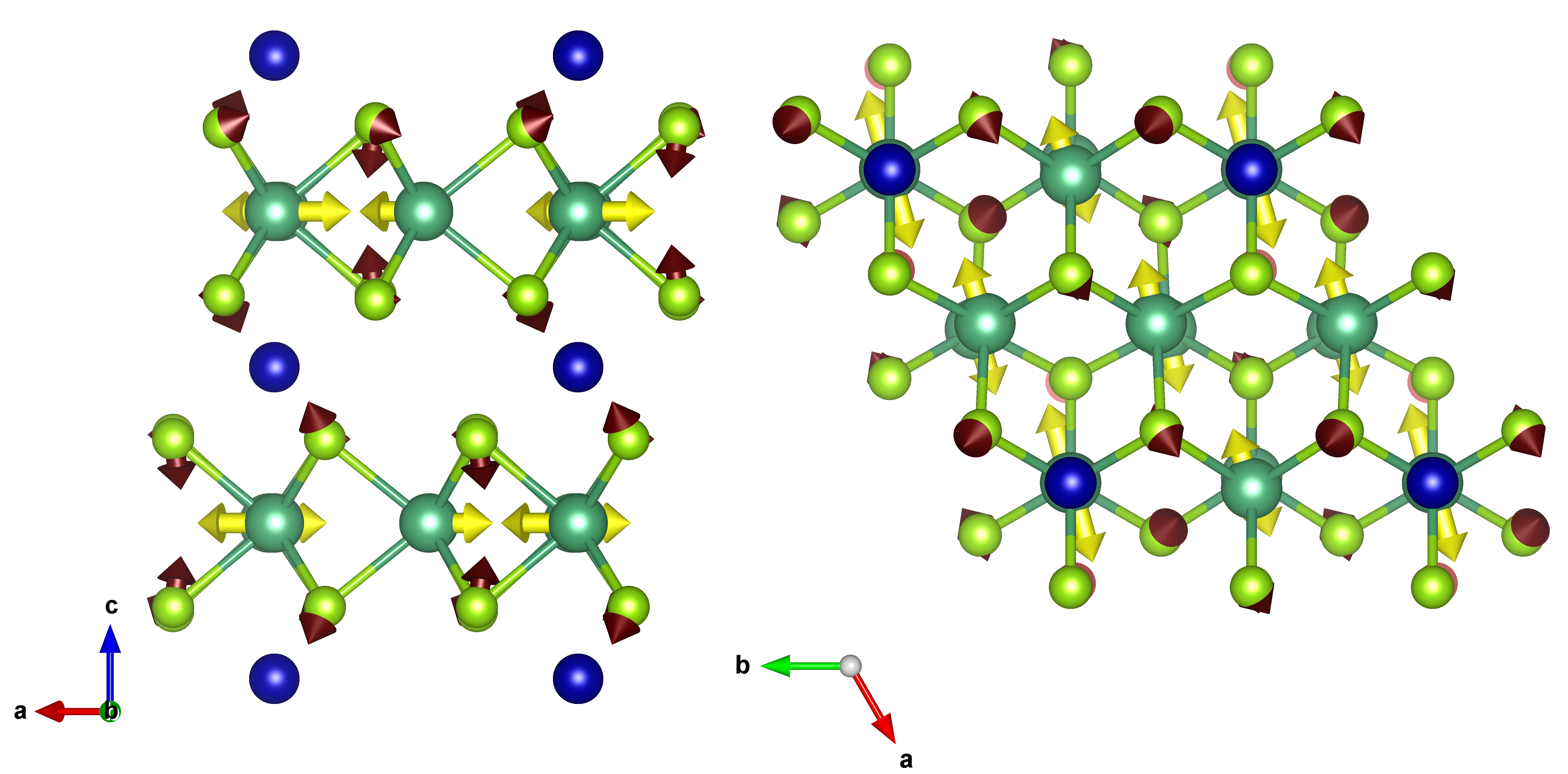}
\captionof{figure}{Two orientations of the 259.17~cm$^{-1}$ ($E^3_{2g}$) phonon mode in Co$_{1/4}$NbSe$_2$.
Left: view along the $b$-axis; right: view along the $c$-axis.
Arrows denote atomic displacement directions.}
\end{minipage}

\end{center}
\clearpage
\begin{center}
\begin{minipage}[t]{0.98\textwidth}
\centering
\captionof{table}{$A_{1g}^3$ phonon mode calculated at 277.41~cm$^{-1}$.}
\label{STab:A1}
\footnotesize 
\setlength{\tabcolsep}{8pt} 
\renewcommand{\arraystretch}{1.1} 
\begin{tabular}{lccc}
\hline
Atom & $\Delta$x & $\Delta$y & $\Delta$z \\
\hline
Co1  &  0.000  & -0.000  &  0.000 \\
Co2  & -0.000  & -0.000  &  0.000 \\
Nb1  & -0.000  &  0.311  &  0.000 \\
Nb2  & -0.000  & -0.000  &  0.000 \\
Nb3  &  0.000  & -0.311  & -0.000 \\
Nb4  &  0.000  &  0.000  & -0.000 \\
Nb5  & -0.269  &  0.155  & -0.000 \\
Nb6  &  0.269  & -0.155  & -0.000 \\
Nb7  & -0.269  & -0.155  &  0.000 \\
Nb8  &  0.269  &  0.155  &  0.000 \\
Se1  &  0.004  &  0.002  &  0.043 \\
Se2  & -0.000  & -0.000  &  0.316 \\
Se3  &  0.000  &  0.005  &  0.043 \\
Se4  &  0.000  & -0.000  & -0.316 \\
Se5  &  0.000  &  0.000  & -0.316 \\
Se6  & -0.000  & -0.005  & -0.043 \\
Se7  & -0.004  & -0.002  & -0.043 \\
Se8  & -0.004  &  0.002  &  0.043 \\
Se9  &  0.000  &  0.000  &  0.316 \\
Se10 &  0.004  & -0.002  & -0.043 \\
Se11 & -0.004  & -0.002  &  0.043 \\
Se12 &  0.004  &  0.002  & -0.043 \\
Se13 & -0.000  & -0.005  &  0.043 \\
Se14 &  0.000  &  0.005  & -0.043 \\
Se15 &  0.004  & -0.002  &  0.043 \\
Se16 & -0.004  &  0.002  & -0.043 \\
\hline
\end{tabular}
\end{minipage}

\vspace{1.0cm}

% ---- Figure Below ----
\begin{minipage}[t]{0.85\textwidth} % slightly narrower for balance
\centering
\includegraphics[width=0.95\linewidth]{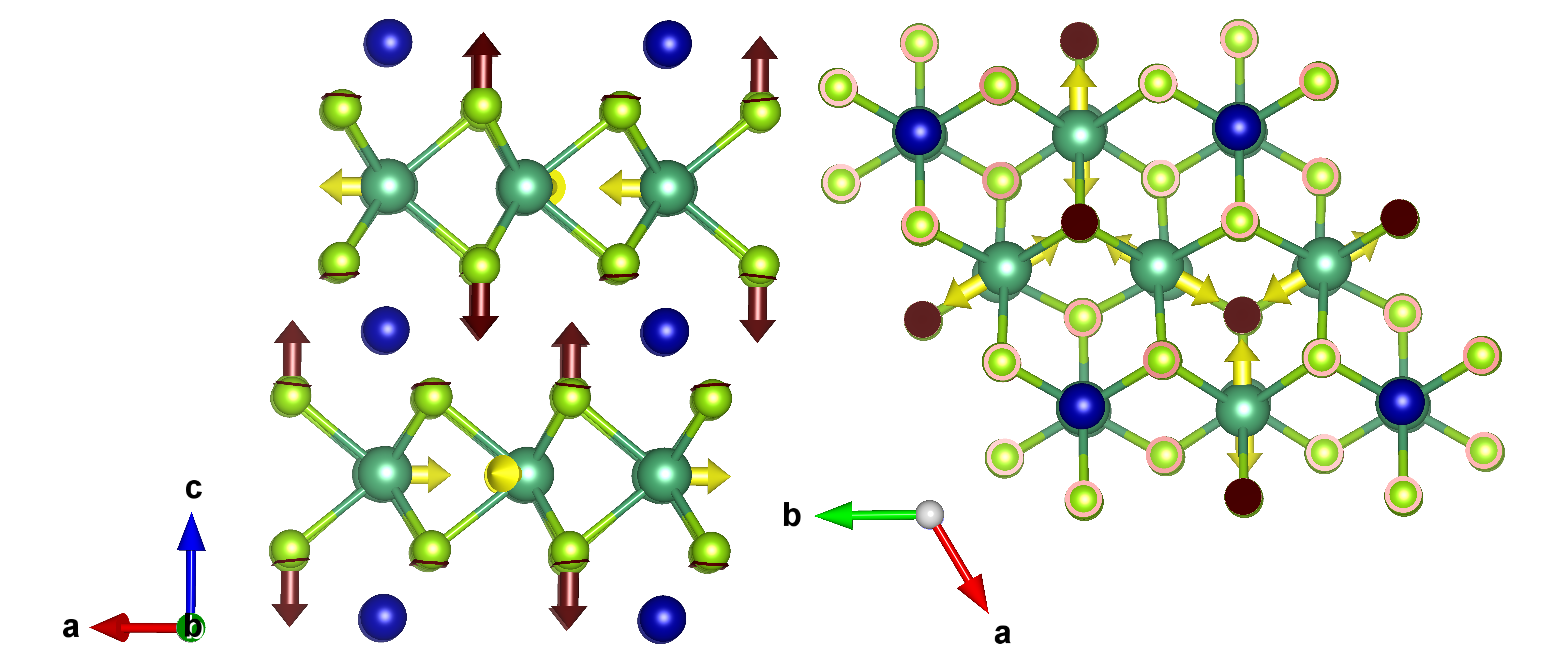}
\captionof{figure}{Two orientations of the 277.41~cm$^{-1}$ ($A^3_{1g}$) phonon mode in Co$_{1/4}$NbSe$_2$. 
Left: view along the $b$-axis; right: view along the $c$-axis. 
Arrows denote atomic displacement directions.}
\end{minipage}
\end{center}
\newpage
\section{Comparison of Raman modes for different magnetic orderings.}

\begin{table}[!htbp]
\caption{Comparison between DFT-calculated Raman mode frequencies (cm$^{-1}$) of Co$_{1/4}$NbSe$_2$ in NM, FM, and AFM phases with 5 K experimental data. The connection to NbSe$_2$ Z.B. and $\Gamma$ point is indicated in the symmetry column.} 
\label{Table:RamanModeComparison}
\begin{ruledtabular}
\begin{tabular}{lcccc}
Symmetry & NM & FM & AFM & Expt (5 K) \\
\hline
$E^{1}_{2g}$(Z.B.) & 87.61  & 107.30 & 98.18  & 102.52 \\
$E^{2}_{2g}$(Z.B.) & 123.66 & 123.99 & 124.17 & 127.16 \\
$A^{1}_{1g}$(Z.B.) & 134.42 & 137.04 & 150.35 & 133.11 \\
$A^{2}_{1g}$($\Gamma$) & 226.37 & 215.47 & 220.36 & 226.18 \\
$E^{3}_{2g}$($\Gamma$) & 252.92 & 254.00 & 259.17 & 257.40 \\
$A^{3}_{1g}$(Z.B.) & 263.05 & 266.64 & 277.41 & 269.83 \\
\end{tabular}
\end{ruledtabular}
\end{table}

\begin{figure*}[t]

\maketitle
\section{Sample Dependence of Anharmonic Behavior}

Figures S9(a)-S9(f) show the temperature dependence of all observed Raman mode frequencies for three independent samples. These data are averaged to obtain the red data set that is used in our analysis in the main text. 

\vspace{0.5em}

\centering

% Row 1
\begin{subfigure}{0.48\textwidth}
    \centering
    \includegraphics[width=\linewidth]{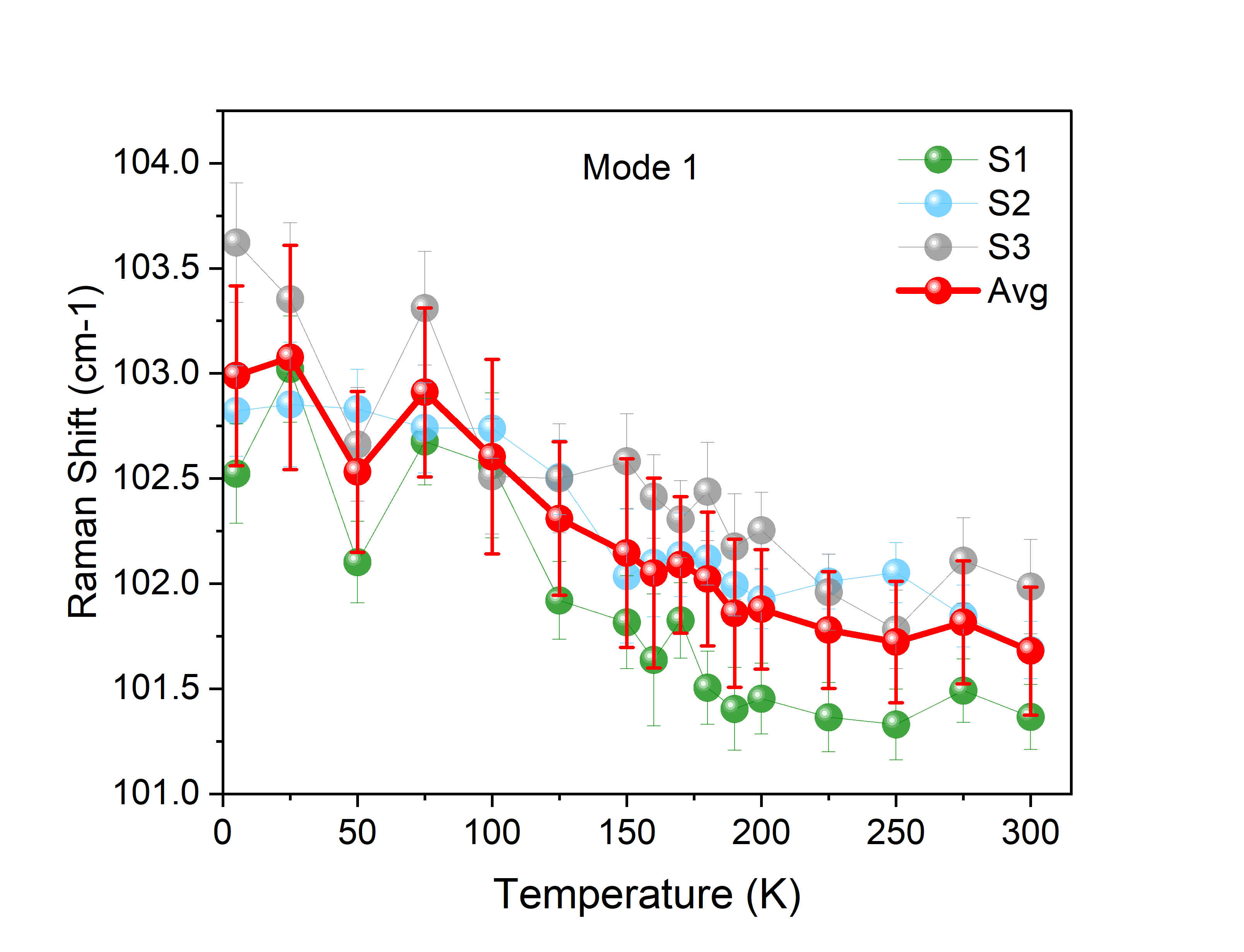}
    \caption{$E_{2g}^1$ (102.52 cm$^{-1}$)}
\end{subfigure}
\hfill
\begin{subfigure}{0.48\textwidth}
    \centering
    \includegraphics[width=\linewidth]{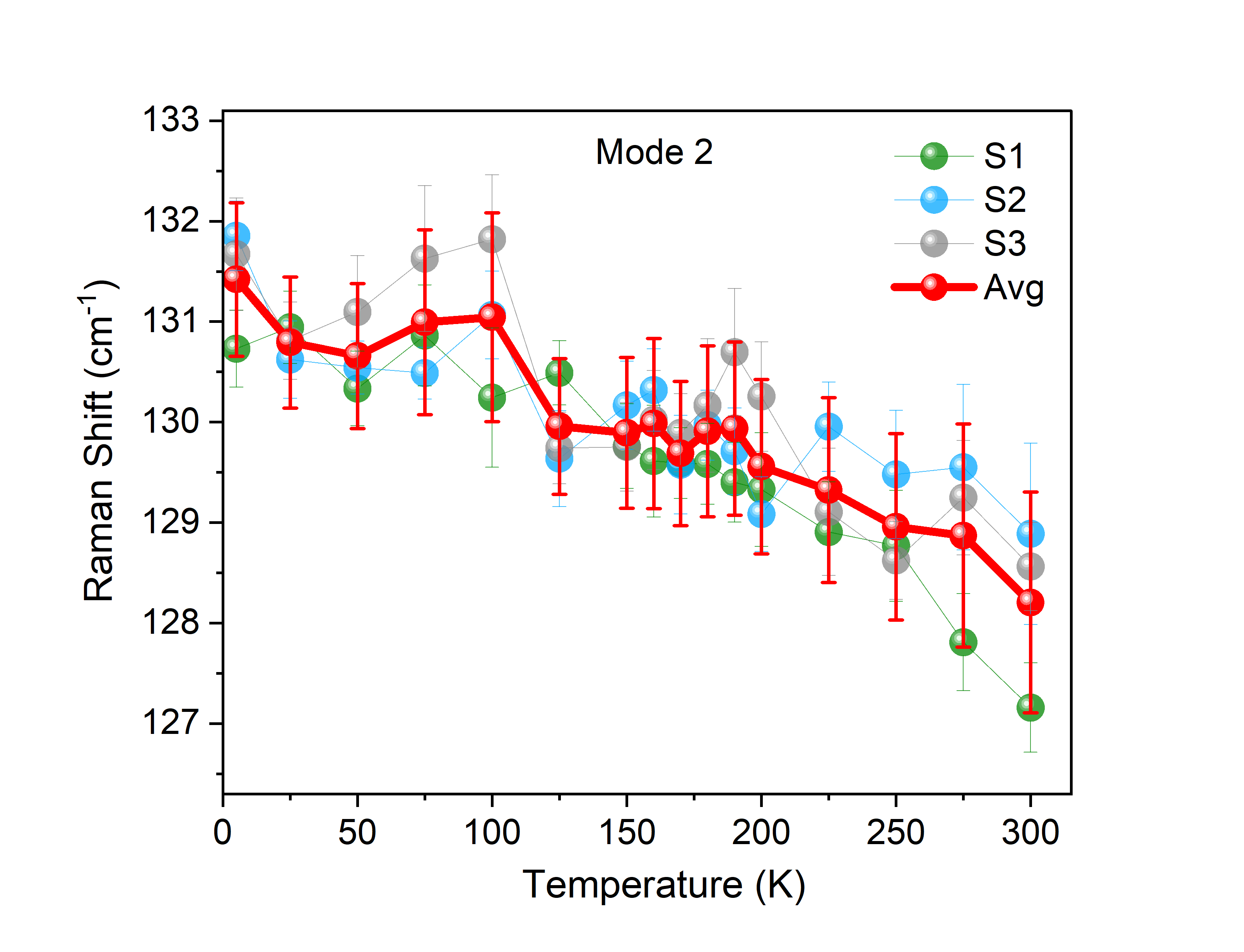}
    \caption{$E_{2g}^2$ (130.73 cm$^{-1}$)}
\end{subfigure}

\vspace{0.5cm}

% Row 2
\begin{subfigure}{0.48\textwidth}
    \centering
    \includegraphics[width=\linewidth]{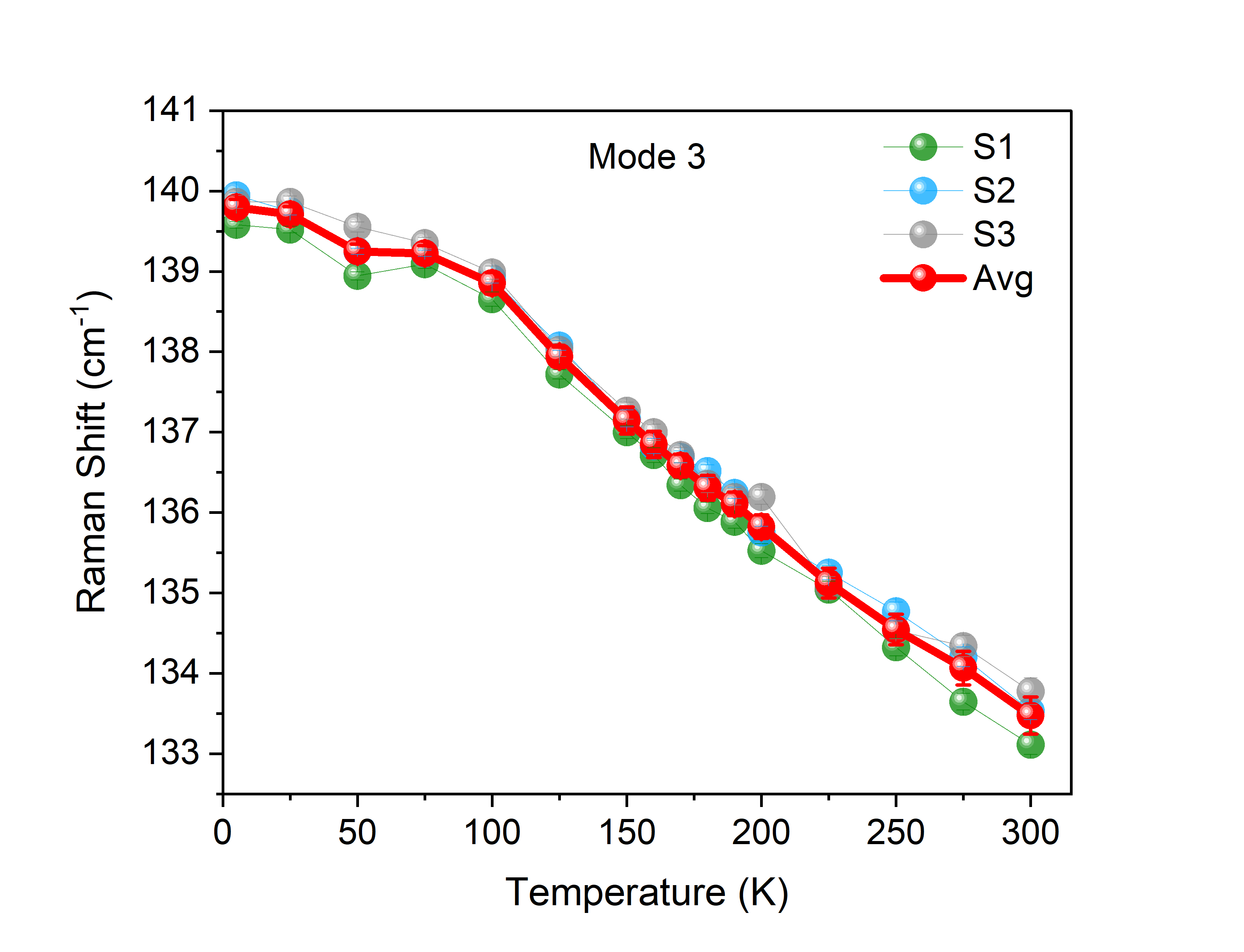}
    \caption{$A_{1g}^1$ (139.58 cm$^{-1}$)}
\end{subfigure}
\hfill
\begin{subfigure}{0.48\textwidth}
    \centering
    \includegraphics[width=\linewidth]{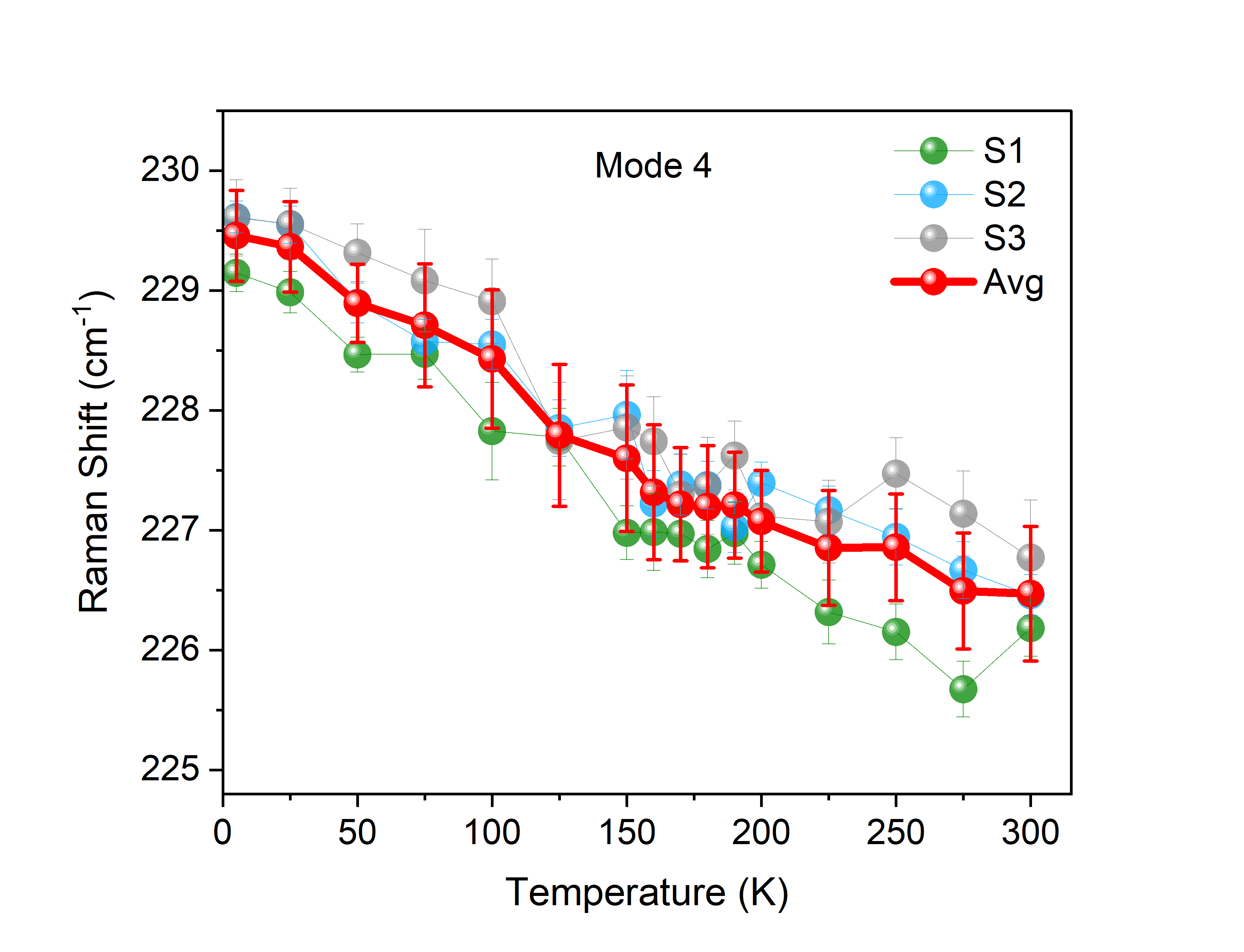}
    \caption{$A_{1g}^2$ (229.15 cm$^{-1}$)}
\end{subfigure}

\vspace{0.5cm}

% Row 3
\begin{subfigure}{0.48\textwidth}
    \centering
    \includegraphics[width=\linewidth]{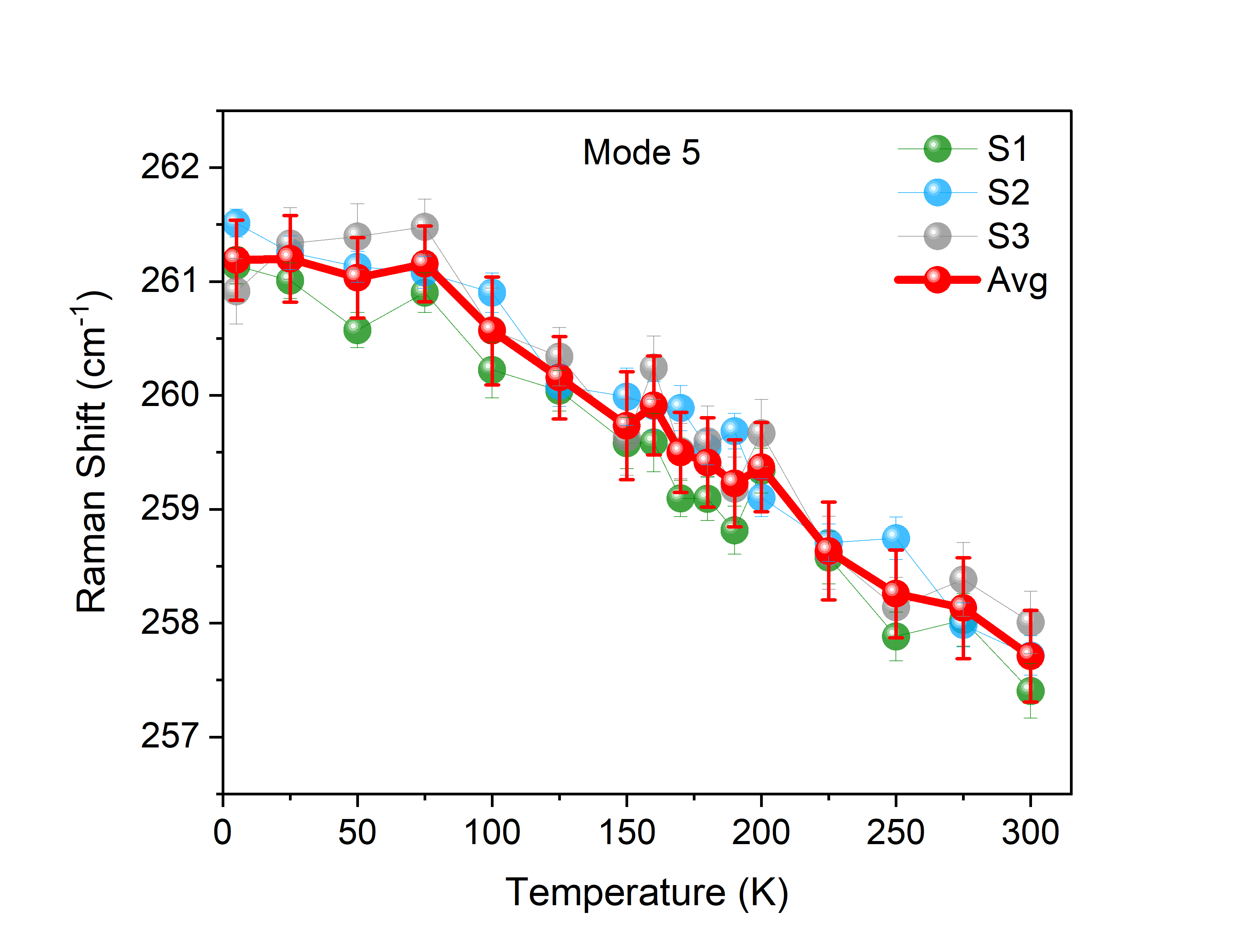}
    \caption{$E_{2g}^3$ (261.14 cm$^{-1}$)}
\end{subfigure}
\hfill
\begin{subfigure}{0.48\textwidth}
    \centering
    \includegraphics[width=\linewidth]{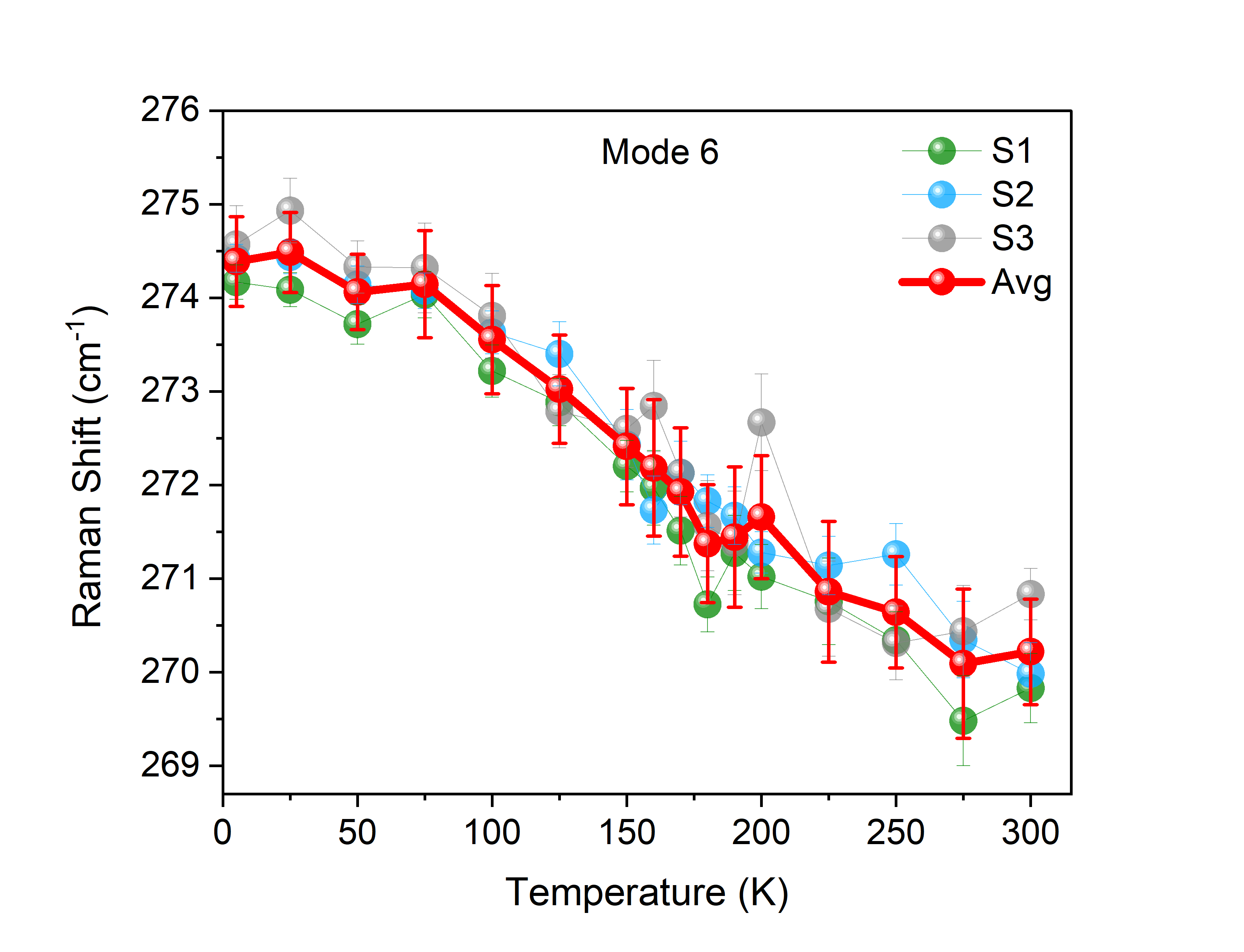}
    \caption{$A_{1g}^3$ (274.17 cm$^{-1}$)}
\end{subfigure}

\caption{Sample dependence of anharmonic phonon behavior. The average value is shown with the red curve.}
\label{fig:S9-S14}

\end{figure*}
\clearpage

\section{Additional Raman Measurements and Sensitivity Analysis}
\begin{figure}[ht]
    \centering
    \includegraphics[width=0.5\textwidth]{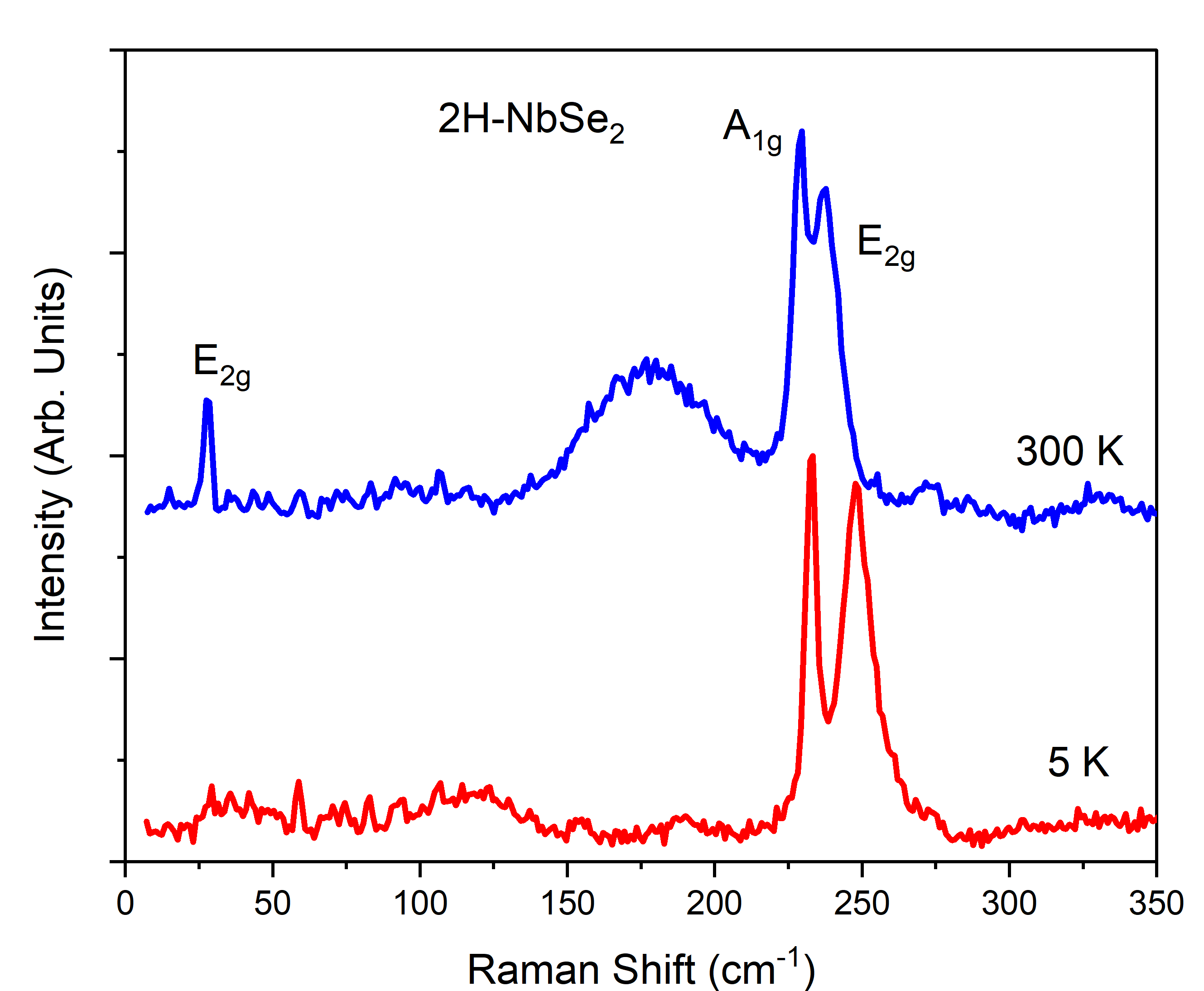}
    \caption{Temperature-dependent low-frequency Raman spectra of bulk 2H-NbSe$_2$ measured at 5 and 300~K. The weak feature near 28~cm$^{-1}$, corresponding to the interlayer shear mode, is clearly resolved at 300~K but becomes substantially weaker at 5~K. The mode lies well above the $\sim$10~cm$^{-1}$ low-frequency cutoff of the Raman setup.}
    \label{fig:S10}
\end{figure}
\clearpage

To examine the sensitivity of the extracted phonon frequency shifts to the choice of fitting range, we repeated the Balkanski analysis by varying the lower bound of the fitting window from 170 to 200~K while keeping the upper bound fixed at 300~K. For each fitting window, the resulting anharmonic fit was extrapolated to 0~K and $\Delta\omega$ was determined according to Eq.~(5). The extracted $\Delta\omega$ values remain comparable across the
different fitting windows, demonstrating that the low-temperature phonon deviations are robust against exclusion of the temperature range immediately above $T_N$. The corresponding fits and extracted $\Delta\omega$ values are shown in Fig.~S11 and summarized in Table~10, respectively.

\vspace{0.7cm}
\begin{figure*}[ht]
    \centering
    \includegraphics[width=0.95\textwidth]{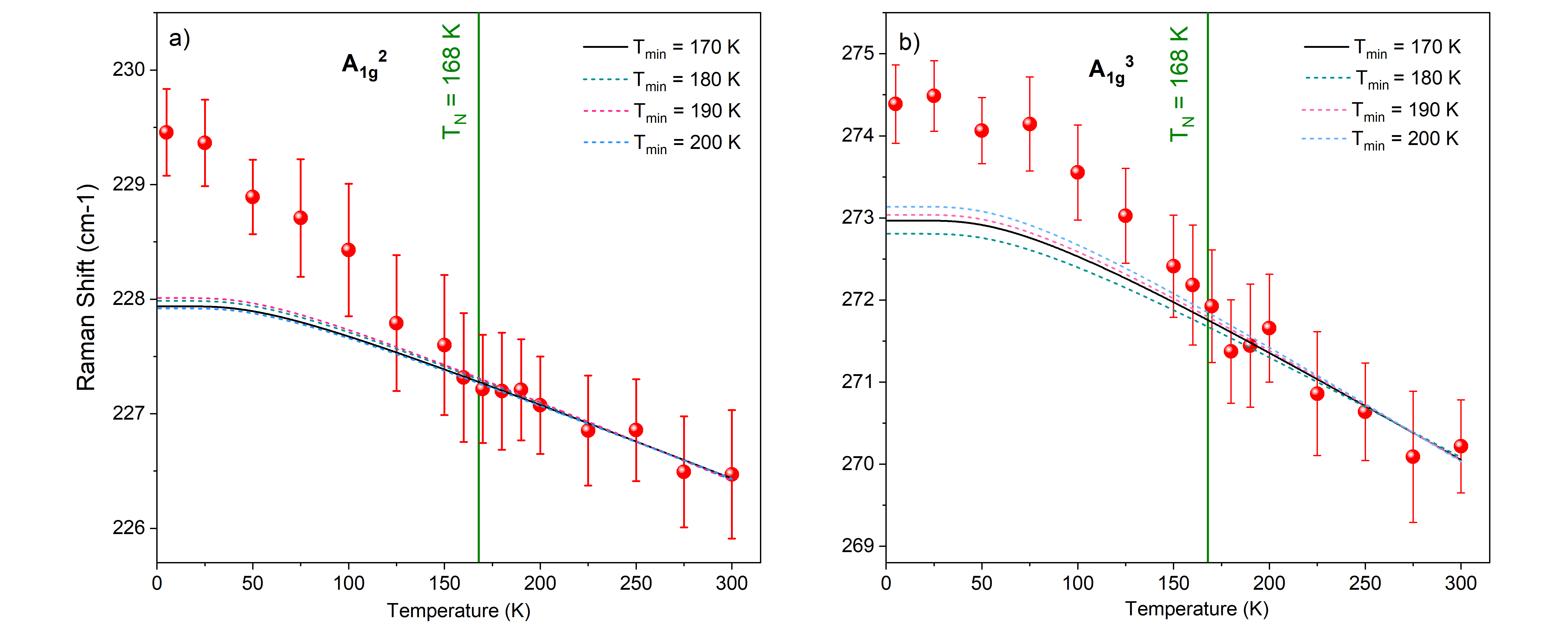}
    \caption{Balkanski fits obtained by varying the lower bound of the fitting window from 170 to 200~K, with the upper bound fixed at 300~K, for all six Raman-active modes of Co$_{1/4}$NbSe$_2$.}
    \label{fig:S11}
\end{figure*}

\vspace{1.2cm}
\begin{table}[ht]
\centering
\large
\renewcommand{\arraystretch}{1.3} % increases vertical spacing
\setlength{\tabcolsep}{12pt}      % increases horizontal spacing between columns

\caption{Sensitivity of the extracted phonon frequency shift
$\Delta\omega$ and spin--phonon coupling constant $\lambda$ to the lower temperature bound $T_{\min}$ used in the Balkanski fit. The upper bound of the fitting range was fixed at 300~K.}
\label{tab:fit_sensitivity}
\vspace{0.5cm}
\begin{tabular}{c c c c}
\hline
Mode & $T_{\min}$ (K) &
$\Delta\omega$ (cm$^{-1}$) &
$\lambda$ (cm$^{-1}$) \\
\hline
\rule{0pt}{2.5ex}
$A_{1g}^{2}$ & 170 & 1.52 & 0.67 \\
              & 180 & 1.47 & 0.65 \\
              & 190 & 1.45 & 0.64 \\
              & 200 & 1.54 & 0.68 \\
\hline
\rule{0pt}{2.5ex}
$A_{1g}^{3}$ & 170 & 1.42 & 0.63 \\
              & 180 & 1.58 & 0.70 \\
              & 190 & 1.35 & 0.60 \\
              & 200 & 1.25 & 0.56 \\
\hline
\end{tabular}
\end{table}